\documentclass[aps,rmp,reprint,amsmath,amssymb,graphicx,longbibliography]{revtex4-1}

\usepackage{bm}
\usepackage{graphicx,amssymb,amsmath,dcolumn,gensymb}

\newcommand{\etal}{\textit{et al}. }

\begin{document}

\title{Astromaterial Science and Nuclear Pasta}
\author{M. E. Caplan}\email{mecaplan@indiana.edu}
\author{C. J. Horowitz}\email{horowit@indiana.edu}
\affiliation{Center for Exploration of Energy and Matter and Department of Physics, Indiana University, Bloomington, IN 47405, USA}

\date{\today{}}

\begin{abstract}
We define `astromaterial science' as the study of materials in astronomical objects that are qualitatively denser than materials on earth.  Astromaterials can have unique properties related to their large density, though they may be organized in ways similar to more conventional materials. 
By analogy to terrestrial materials, we divide our study of astromaterials into hard and soft and discuss one example of each. The hard astromaterial discussed here is a crystalline lattice, such as the Coulomb crystals in the interior of cold white dwarfs and in the crust of neutron stars, while the soft astromaterial is nuclear pasta found in the inner crusts of neutron stars. 
In particular, we discuss how molecular dynamics simulations have been used to calculate the properties of astromaterials to interpret observations of white dwarfs and neutron stars. 
Coulomb crystals are studied to understand how compact stars freeze.  Their incredible strength may make crust ``mountains'' on rotating neutron stars a source for gravitational waves that the Laser Interferometer Gravitational-Wave Observatory (LIGO) may detect.  Nuclear pasta is expected near the base of the neutron star crust at densities of $10^{14}$ g/cm$^3$.  Competition between nuclear attraction and Coulomb repulsion rearranges neutrons and protons into complex non-spherical shapes such as sheets (lasagna) or tubes (spaghetti). Semi-classical molecular dynamics simulations of nuclear pasta have been used to study these phases and calculate their transport properties such as neutrino opacity, thermal conductivity, and electrical conductivity. Observations of neutron stars may be sensitive to these properties, and can be be used to interpret observations of supernova neutrinos, magnetic field decay, and crust cooling of accreting neutron stars.  We end by comparing nuclear pasta shapes with some similar shapes seen in biological systems.
\end{abstract}


\maketitle

\tableofcontents{}
\section{Introduction}
\label{sec.intro}
The heavens contain a variety of materials that range from conventional to extraordinary and extreme.   Solar system bodies such as the moon, asteroids, and terrestrial planets contain rocks that are similar to rocks on earth, however, observations that 40 Eridani B \cite{Adams1914} and Sirius B \cite{Adams1915} are dim and hot stars led to a dramatic change. These white dwarfs are much denser than anything on earth and have central densities more than a million times that of water, and thus may be made of materials unlike any we can produce in the laboratory.   

The boundaries of the field of material science are difficult to define; perhaps a `material' may simply be described as matter that is not fluid. For this colloquium, we define `astromaterial science' as the study of materials in astronomical objects that are qualitatively denser than terrestrial materials.  Astromaterials can have unique properties related to their density, such as extraordinary mechanical strength, though they may be organized in ways similar to more conventional materials.  

Advances in computational material science may be particularly useful for astromaterial science because astromaterials typically are present under extreme conditions, such as very high pressure, making them inaccessible to laboratory experiments on earth.  Although we are not aware of any previous reference to astromaterial science, there are a number of works discussing related concepts, such as stellar metallurgy \cite{Kobyakov2014,Ball2014}. Chamel also studied metallurgical aspects of the neutron star crust by showing parallels between band structure of neutrons in the crust and electrons in metals \cite{CHAMEL2005109}.

Often materials are classified as hard or soft.  For example, a crystalline solid is a hard material.  Coulomb crystals in the interior of cold white dwarfs and in neutron star crusts are examples of hard astromaterials.  In contrast, a soft material is neither a liquid nor a crystalline solid.  Many soft materials are common in everyday life, such as gels and glasses, liquid crystals, and biological membranes which may involve the self-assembly of microscopic components into much larger complex shapes.

By analogy to self-assembly in terrestrial soft matter, we consider nuclear pasta a soft astromaterial.  Pasta is expected near the base of the neutron star crust at densities of  $10^{14}$ g/cm$^3$, 100 trillion times that of water.  This density is so high that individual atomic nuclei start to touch.  When this happens, competition between nuclear attraction and Coulomb repulsion rearranges the neutrons and protons from nuclei into complex non-spherical shapes such as flat sheets (lasagna) or thin tubes (spaghetti). These shapes may influence transport and elastic properties of nuclear pasta, making pasta relevant for interpreting a number of neutron star observations.

This colloquium discusses one hard astromaterial, Coulomb crystals, and one soft astromaterial, nuclear pasta.  First, we review the structure of white dwarfs and neutron stars in Section \ref{sec.astrosolid}.  Next, Section \ref{sec.hard} discusses Coulomb crystals in the interior of cold white dwarfs and in neutron star crusts.  We begin by considering how stars freeze, including discussing the limited observations of solidifying white dwarfs and the cooling of neutron star crusts that occurs after accretion heating from a companion star.  We end this section with an application of hard astromaterials to the radiation of continuous gravitational waves from ``mountains'' on rotating neutron stars.  

We describe large-scale molecular dynamics simulations of the breaking stress of Coulomb crystals.  On earth the breaking stress of rock is the limiting factor in how large a mountain can be.  Mountains can also form on neutron stars, although because of the immense gravity their heights are measured in centimeters instead of kilometers. Similar to terrestrial rock, the breaking stress of neutron star crust is the limiting factor to their size.   Molecular dynamics simulations suggest the breaking stress of neutron star crust is enormous, some ten billion times larger than steel.  This strong crust can support mountains large enough to emit detectable gravitational waves as the star spins.


Section \ref{sec.soft} is on nuclear pasta.  After a historical review we describe semi-classical molecular dynamics simulations of nuclear pasta and use them to illustrate some of the complex shapes that are possible.  We discuss transport properties of nuclear pasta including shear viscosity, thermal conductivity, electrical conductivity, and neutrino opacity.  We also discuss topological defects in the pasta and observations of magnetic field decay and crust cooling of neutron stars that may be sensitive to pasta properties.  Finally, we compare nuclear pasta shapes to some similar shapes seen in biological systems made of phospholipids.  

We summarize and offer a short perspective in Sec. \ref{conclusions}.

\section{Compact stars and Astro-Solids}
\label{sec.astrosolid}
Stars are made of hot plasmas.  Nevertheless, compact stars such as white dwarfs (WDs) and neutron stars (NSs) are so dense that the plasma can crystallize. We begin by reviewing the structure of cold WDs and NSs.

\begin{figure}[ht!]
\centering
\includegraphics[width=0.49\textwidth]{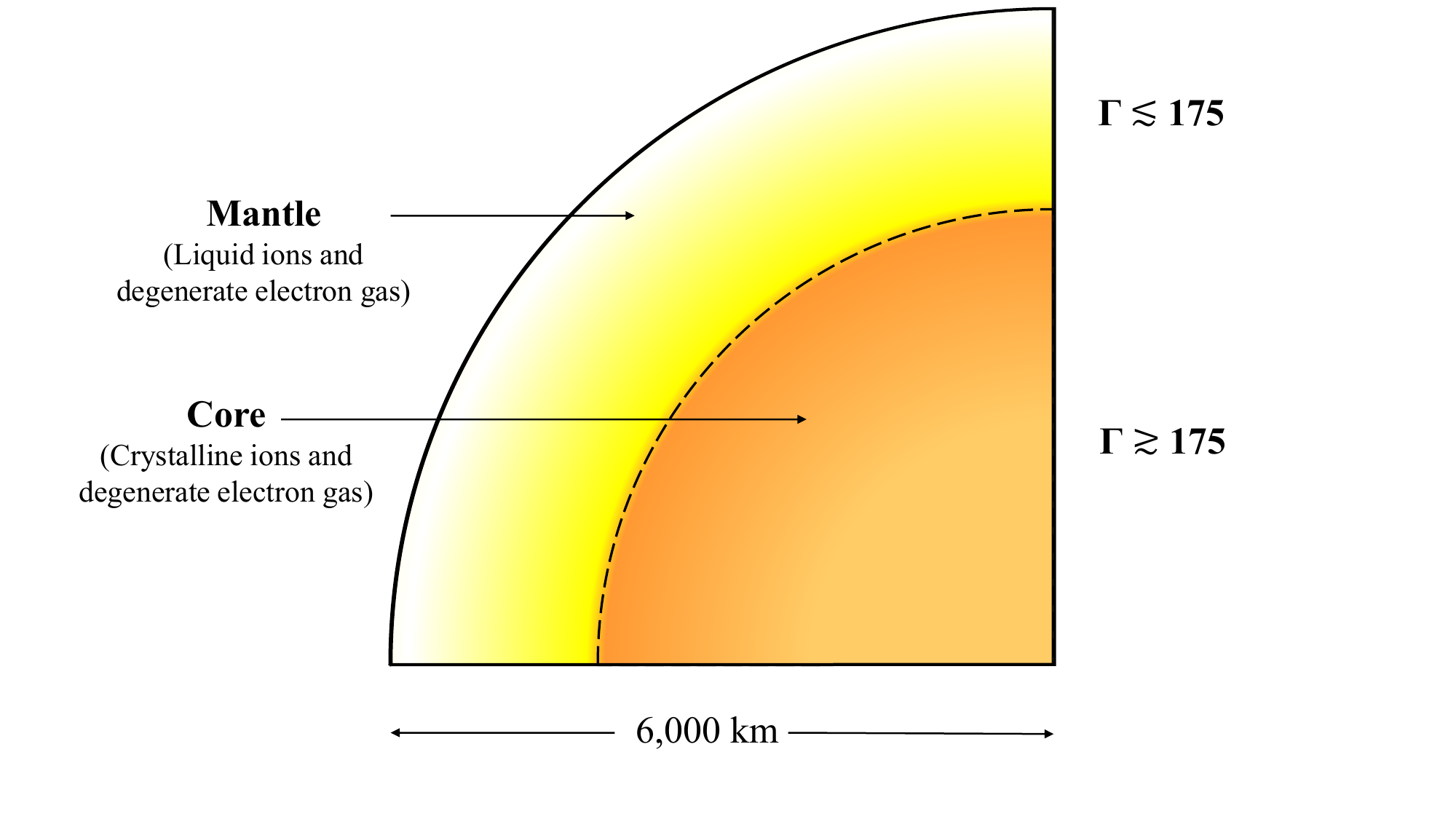}
\caption{\label{fig:WD_PieChart}  Cross section of a cooling white dwarf star. The Coulomb crystals described in Sec. \ref{sec.hard} are found in the solid core, while the mantle is liquid.  The Coulomb parameter $\Gamma$ describes the ratio of a typical Coulomb energy between ions to the thermal energy $kT$, see Eq. \ref{eq:gamma}. } 
\end{figure}

Figure \ref{fig:WD_PieChart} shows the expected structure of a cold WD.  There is a liquid mantle composed of ions and a degenerate electron gas.  The Coulomb parameter $\Gamma$ describes the ratio of a Coulomb energy $Z^2e^2/a$ between ions (of charge $Z$), to the thermal energy $kT$, see below.  Here $a$ is a typical distance between ions.  The density increases by several orders of magnitude as one approaches the center of the star, so $a$ decreases with depth\footnote{The central density of a WD could be as high as $10^9$ g/cm$^3$}. This increases the Coulomb energy and $\Gamma$ (in the approximately isothermal interior) and causes the material to crystallize.  Thus, the star has a solid core surrounded by a liquid mantle.  As the star continues to cool the size of the solid core will grow with time. WD crystallization is discussed in more detail in Sec. \ref{subsec.freeze}.

\begin{figure}[ht!]
\centering
\includegraphics[width=0.49\textwidth]{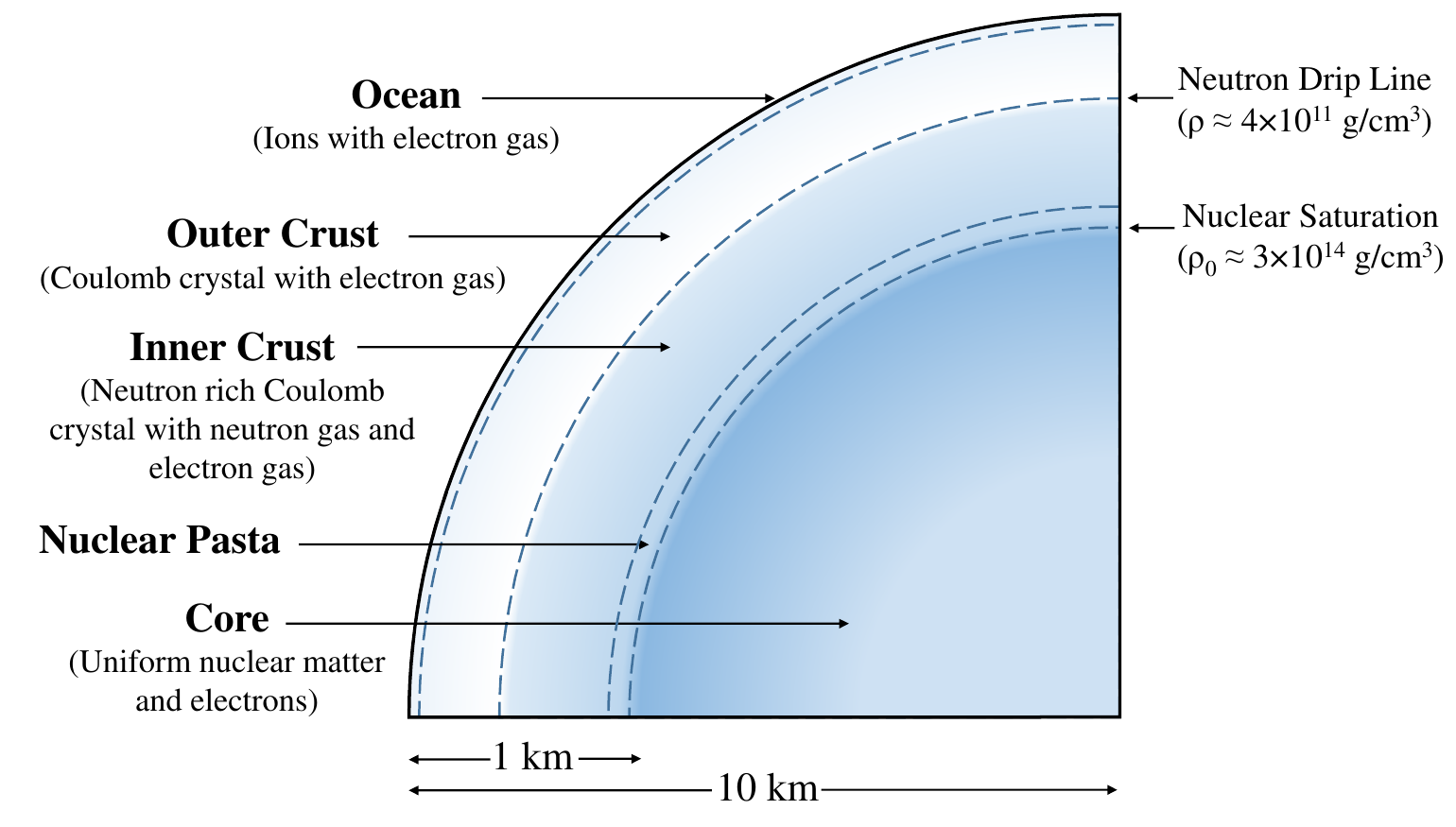}
\caption{\label{fig:NS_PieChart}  Cross section of a neutron star. The Coulomb crystals described in Sec. \ref{sec.hard} are found in the inner and outer crusts, while the nuclear pasta, described in Sec. \ref{sec.soft}, may be found at the base of the inner crust. } 
\end{figure}

Figure \ref{fig:NS_PieChart} shows the expected structure of a NS.  Warm material accreting onto a NS can form a liquid ocean that is similar to the mantle of a WD.  This material is then buried as more accreted matter arrives, so the density and $\Gamma$ increase until this material reaches the bottom of the ocean where it freezes to form the outer crust.  The outer crust contains crystalline ions and degenerate electrons.  It is similar to the solid core of a cold WD.   As the density continues to increase, the rising electron Fermi energy drives electron capture reactions $e+p\rightarrow n +\nu_e$.  This makes the ions in the outer crust more and more neutron rich.  When the density reaches about $10^{11}$ g/cm$^3$ the ions become so neutron rich that neutrons ``drip'' out of these nuclei.  This forms the inner crust of a NS which is composed of a gas of nearly free neutrons, crystalline ions, and a very degenerate gas of relativistic electrons.

As one approaches nuclear saturation density $n_0\approx 3\times 10^{14}$ g/cm$^3$ (the density inside a nucleus) individual nuclei approach one another and begin to touch.  At this point competition between short range nuclear attraction from the strong force and long range Coulomb repulsion can rearrange nearly spherical nuclei into tubes (spaghetti), flat sheets (lasagna) or other exotic shapes that are known as nuclear pasta  \cite{1984PThPh..71..320H, PhysRevLett.50.2066}.  These nuclear pasta phases are expected at the base of the inner crust and we discuss them in Sec. \ref{sec.soft}.  Finally at densities above $n_0$, in the core of a NS, one expects a uniform liquid of neutron rich matter.  Here quantum zero point motion keeps the liquid from crystallizing.   We now discuss how these solid phases form in WDs and accreting NSs.


\section{Coulomb crystals as hard astromaterials}
\label{sec.hard}
How do WDs and NSs freeze? 
At WD densities and beyond, material is pressure ionized.  The electrons form a degenerate Fermi gas while the ions interact via Coulomb potentials $V(r)$ that are only slightly screened by the electrons,
\begin{equation}
V(r_{ij})=\frac{Z_i Z_j e^2}{r_{ij}} \exp(-r_{ij}/\lambda)\, .
\label{eq.V}
\end{equation}  
Here $r_{ij}$ is the distance between ions of charge $Z_i$ and $Z_j$.  The electron Fermi energy is typically the largest energy in the problem and limits the electron polarizability.   Therefore electrons are not treated explicitly.   Instead electrons are included by adding a screening factor to the Coulomb interaction. This screening factor is taken to be the Thomas-Fermi screening length $\lambda$, which is given by 
\begin{equation}
\lambda=\frac{\pi^{1/2}}{(4\alpha k_F)^{1/2}(k_F^2+m_e^2)^{1/4}}\sim\frac{1}{2k_F}\sqrt{\frac{\pi}{\alpha}}
\label{eq.lambda}
\end{equation}
for non-interacting relativistic electrons.   The screening length is a function of the electron mass $m_e$ and its Fermi momentum $k_F=(3\pi^2n_e)^{1/3}$.  The electron density $n_e$ is equal to the ion charge density, $n_e=\langle Z\rangle n$, where $n$ is the ion density and $\langle Z\rangle$ is the average charge.

The system can be characterized by a Coulomb parameter $\Gamma$ describing the ratio of a typical Coulomb energy to the thermal energy.  For a one component plasma (OCP), where all of the ions have charge $Z$ at a temperature $T$, we have,
\begin{equation}
\Gamma=\frac{Z^2e^2}{akT}\, ,
\label{eq:gamma}
\end{equation}
with $a=(3/4\pi n)^{1/3}$.

In many cases the OCP is a useful model, though in general one may be interested in a mixture of ions of different charges. In the case of a multi-component plasma (MCP) species are first treated individually by calculating the ion sphere radius $a_{i}=(3 Z_{i}/4\pi n_e)^{1/3}$ for each species $i$. This gives $\Gamma_{i}={Z_{i}^2e^2}/(a_{i}kT)$ for each species so that averaging over all ions gives $\Gamma$ for the mixture,
\begin{equation}
\Gamma = \frac{ \langle Z^{5/3} \rangle e^{2} } {kT} \left [  \frac{4 \pi n_e}{3} \right ]^{1/3}\ .
\label{eq:gammaMCP}
\end{equation}


The determination that the OCP freezes at $\Gamma\approx 175$ has a long history starting with  the Monte Carlo simulations of Brush, Sahlin and Teller in 1966 \cite{jchemphys1966}.  Since then during the 1960s to 1990s there have been many works including \cite{Pollock73,Slattery80} who determined the melting $\Gamma$ with improved accuracy, while \cite{Loumos73,Ichimaru88} studied freezing of Carbon Oxygen mixtures, and \cite{Jones96} included quantum effects using path integral Monte Carlo. 

In contrast to the OCP, the MCP does not have a fixed freezing temperature; rather, it depends on the exact composition. Much work has been done to calculate $\Gamma$ for the MCP due to its importance in phase separation in the neutron star ocean. Recently, Medin and Cumming \cite{Medin2010} have developed a semi-analytical method for calculating phase diagrams of arbitrary mixtures by generalizing early work on one-, two-, and three-component plasmas.  This method has since been applied to many different mixtures of rp-process ash to find $\Gamma$ \cite{Mckinven:2016zkg}. 
 

We now introduce molecular dynamics (MD) simulations that are similar to the Monte Carlo simulations that were originally used.  We will discuss a number of MD simulation results later in this review.  A simulation involves $N$ particles that are at positions ${\bf r}_i(t)$ with velocities ${\bf v}_i(t)$ in a simulation volume that typically has periodic boundary conditions.   The force on each particle, from the surrounding particles, is calculated using the chosen force law (for example, the negative gradient of Eq. \ref{eq.V}).  This is often the most time consuming step.  Then, Newton's laws are used to calculate the new ${\bf r}_i(t+\Delta t)$ and ${\bf v}_i(t+\Delta t)$ at simulation time $t+\Delta t$.  Here $\Delta t$ is a small MD time step.\footnote{Many algorithms exist for integrating Newton's laws efficiently and accurately; popular algorithms include Velocity-Verlet \cite{PhysRev.159.98} and predictor-corrector.} Starting from some initial configuration one repeats this procedure for many MD time steps until the system comes into equilibrium. Then one can ``measure" quantities of interest such as the potential energy by simply evaluating the quantity for the positions ${\bf r}_i(t)$ at time $t$.  Ensemble averages of observables can be calculated by making many measurements at different simulation times and forming a time average.   Note that MD simulations ignore quantum effects on the ions.  These were considered for binary Coulomb crystals by \cite{doi:10.1063/1.4930215}.

\subsection{How stars freeze: Coulomb crystals in White Dwarfs}
\label{subsec.freeze}
Observations of cooling white dwarf (WD) stars provide important information on the ages and evolution of stellar systems \cite{cosmochron, garcia-berro, renedo, salaris1}.  The interior of a WD is a Coulomb plasma of ions and a degenerate electron gas.  As the star cools this plasma crystallizes.  Note that WDs freeze from the center outward, because of the higher central densities, while NS are expected to form a solid crust over a liquid core.\footnote{Note that the NS should not be thought of as an `inverted WD' with a liquid interior and solid exterior. The liquid core of a NS is fundamentally different from the liquid ocean of a WD. The outer regions of both NS and WD are liquid plasmas of ions, which we call the `mantle' of the WD and the `ocean' of the NS. At high pressure, this plasma freezes, forming the `core' of the WD or the `crust' of the NS. The core of the NS is a superfluid of nucleons at nuclear density, and may more aptly be thought of as the next phase up a ladder of increasing density.} 

This crystallization can delay WD cooling as the latent heat is radiated away, see for example ref. \cite{salaris}.  Winget et al. observed effects from the latent heat of crystallization on the luminosity function of WDs in the globular cluster NGC 6397 \cite{winget}.   Winget et al. suggested the melting temperature of the carbon and oxygen mixtures expected in these WD cores is close to the melting temperature of pure carbon.  This is in agreement with our molecular dynamics simulations  \cite{WD_PRL}.   In addition, asteroseismology provides an alternative way to study crystallization in WD, see for example \cite{astroseismology}.  

Carbon-Oxygen white dwarfs can also contain some Neon.  The liquid-solid phase diagram of the three component C, O, Ne system has now been determined by direct MD simulation \cite{PhysRevE.86.066413}.   The optimal structures of three component He-C-O and C-O-Ne white dwarfs were determined using a genetic optimization technique \cite{Engstrom}.  This technique was originally developed to optimize the structure of conventional materials.   Finally, updated evolutionary sequences for Hydrogen-deficient white dwarfs have been created using the liquid-solid phase diagram from MD simulations \cite{0004-637X-839-1-11}.

\subsection{Crystallization on accreting neutron stars}
\label{subsec.accretingNS}

We next consider freezing on neutron stars (NS). 

The crust of an isolated NS is often well described by the OCP. Being that the NS is made from the core collapse of a supernova, the composition of the crust, for low densities, is commonly assumed to be $^{56}$Fe. With increasing depth, and thus increasing Fermi energy, electron captures occur that drive the matter to be more neutron rich. Successive electron capture layers produce specific neutron rich isotopes at specific depths. As the density gets higher in the inner crust, pycnonuclear reactions can cause nuclei to fuse. Although the general idea is quite simple, the exact isotopes present are not fully understood, as they depend on the exact binding energies and masses of neutron rich nuclei that are not well studied. 

In contrast to the isolated star, a NS in a binary system with a conventional star can accrete matter from its companion, replacing the crust with a complex mixture.  Hydrogen rich material falling on the NS often undergoes explosive nuclear burning where protons are rapidly captured by seed nuclei (rp-process nucleosynthesis) to build up heavier nuclei with mass numbers $A$ up to $A\approx 107$ \cite{rpash, rpash2}. This upper limit is due to a closed cycle of burning known as the SnSbTe cycle, which limits rp-process nucleosynthesis to $Z \leq 52$ \cite{rpash}.

As this rp-process material is buried by further accretion, the rising electron Fermi energy induces electron capture to produce a range of neutron rich nuclei from O to approximately Se \cite{rpash3}. This material freezes when the density exceeds about $10^{10}$ g/cm$^3$, so that $\Gamma >$ 175.  Our large scale MD simulations of how this complex rp process ash freezes \cite{phasesep} find that chemical separation takes place so that the liquid phase is greatly enriched in low atomic number $Z$ elements, while the newly formed solid crust is enriched in high $Z$ elements.   This chemical separation has now been studied for a variety of compositions \cite{Mckinven:2016zkg} and can lead to convection in the ocean \cite{Medin:2014zfa}, and this convection could be observable in the X-ray light curve \cite{2041-8205-783-1-L3}.

Nuclear reactions typically produce a complex composition with a variety of different $Z$ ions. In terrestrial materials, the presences of `impurities' with different $Z$ can disrupt the lattice structure of a material. However, MD simulations suggest that a regular crystal lattice forms even though large numbers of impurity ions, with different $Z$, may be present.  This regular crystal should have a high thermal conductivity. 
Early work suggested that the lattice may be amorphous, though we do not find an amorphous solid that would have a low thermal conductivity \cite{soliddiffusion}. Recent X-ray observations of neutron stars find that the crust cools quickly when heating from extended periods of accretion abruptly stops \cite{crustcooling,crustcooling1,crustcooling2}.  This is consistent with MD simulations, and strongly favors a crystalline crust over an amorphous solid that would cool more slowly \cite{crustcooling3,crustcooling4}.  Furthermore, this rapid crust cooling suggests that the number of impurities is not too high.  Indeed, recent observations of KS 1731-260 support a low impurity parameter \cite{0004-637X-833-2-186}.  Otherwise, electron-impurity scattering could limit the thermal conductivity.  One possibility is that nuclear reactions burn away impurities as they get buried into the inner crust at densities of order $10^{11}$ g/cm$^3$.


\subsection{Gravitational waves and the strength of astromaterials}
\label{subsec.GW}
Albert Einstein, a century ago, predicted that accelerating masses will produce oscillations of space-time known as gravitational waves (GWs).  
After decades of work, GWs were finally observed by the Laser Interferometer Gravitational-Wave Observatory (LIGO) \cite{PhysRevLett.116.061102}.  LIGO involves two 4 kilometer detectors that are sensitive to GWs at frequencies from about 10 Hz to a few thousand Hz.  

In principle, one needs large masses with large accelerations in order to produce a detectable GW signal. For example, the first GWs detected were 
from the merger of two $\approx 30 M_{\odot}$ black holes orbiting at relativistic speeds. 

However, black holes are not necessary for producing GWs. In fact, any isolated compact body can produce GWs. All one needs is to put a mass on a stick and shake vigorously, though for the GWs to be detectable the mass must be very large and the stick very strong.  Furthermore, the system must be compact, and hence very dense, in order to radiate at the relatively high LIGO frequencies. Therefore, if we seek a compact GW source then we should look for `strong sticks' which may be made of extraordinarily strong astromaterials.

Such a compact body would be a continuous source of GWs, that may be detectable by LIGO.  A surprisingly efficient continuous GW source is a large mountain on a rapidly rotating NS.  This can be a strong source because a large mass undergoes large accelerations.  The limiting variable may be the strength of the NS crust (strong stick) that holds up the mountain. Here, one may be interested not just in the strength of this material or that material, but what is the strength of the strongest possible material?

We performed large scale MD simulations of the breaking stress (strength) of neutron star crust \cite{crustbreaking,chugunov,lowmassNS}.  This determines how large a neutron star mountain can be before it collapses under the star's extreme gravity.  During the simulation a sample is strained by moving top and bottom layers in opposite directions or by deforming the shape of the simulation volume \cite{crustbreaking}.  These simulations explored the effects of defects, impurities, and grain boundaries on the breaking stress.  
The system may start to break along grain boundaries.  However the large pressure holds the microcrystals together and the system does not fail until large regions are deformed.     
We find that neutron star crust is very strong because the high pressure prevents the formation of voids or fractures and because the long range Coulomb interactions ensure many redundant ``bonds'' between planes of ions.  Neutron star crust is the strongest material known, according to our simulations.  The breaking stress is 10 billion times larger than that for steel.  This is very promising for GW searches because it shows that large mountains are possible, and these could produce detectable signals.  

Our simulations have a breaking strain, fractional deformation when the crust fails, of order 0.1.  This can support significant mountains with a large mass asymmetry and an ellipticity $\epsilon$ as large as $10^{-6}$ to $10^{-5}$.  The ellipticity is the fractional difference in moments of inertia $\epsilon=(I_1-I_2)/I_3$ where  $I_1$, $I_2$ and $I_3$ are the three principle moments of inertia of a neutron star.

A rotating asymmetric neutron star efficiently radiates gravitational waves.  The gravitational wave strain $h_0$ is the fractional change in length of LIGO's arms from the passage of a GW.   The gravitational wave strain due to a rotating NS a distance $d$ away is
\begin{equation}
h_0=\frac{4\pi^2G I_3f^2}{dc^4} \epsilon \, .
\label{eq.h0}
\end{equation}
Here $G$ is Newton's constant, $c$ is the speed of light and $f$ is the frequency of GW radiation.  This frequency is twice the star's rotational frequency $f=2\omega$.  Equation \ref{eq.h0} is very simple and involves largely known quantities such as $I_3$.  Observations of, or upper limits on, $h_0$ can be used to set observational limits on $\epsilon$.   Extensive searches for GW from known pulsars have now been performed \cite{2014ApJ...785..119A,0004-637X-839-1-12}.  No sources were detected but several upper limits on $\epsilon$ were obtained, which in the most sensitive cases were $\epsilon<10^{-8}$.  Neutron star crust can support large mountains yielding ellipticities larger than this.  However, the observations show that, for these stars at least, such large mountains did not form.  Processes that build mountains on NS are largely unknown; mountain building could be associated with accretion where matter may accumulate asymmetrically because of magnetic fields or temperature gradients.  

Neutron stars that accrete matter from a companion may be interesting sources of GW.  No neutron star has been observed to spin faster than about 700 Hz.  This could be because the spin up torque from accretion is balanced by a spin down torque from GW radiation \cite{1538-4357-501-1-L89,0004-637X-510-2-846}.  If this GW torque is from crust mountains, this only requires a relatively modest $\epsilon\approx 10^{-8}$ \cite{Patruno:2017oum}.  Searches for GW from the accreting systems Scorpius X-1 \cite{Abbott:2017mwl} and XTE J1751-305 have now been reported \cite{PhysRevD.95.042005}.

There are several other searches for continuous GW.  For example, a computationally expensive all sky search for GW from unknown neutron stars was sensitive to rapidly spinning stars with $\epsilon=10^{-6}$ out to distances of 1 kpc \cite{arXiv:1605.03233}.  Note that neutron stars are small and can be difficult to observe.  There are undoubtedly vastly more neutron stars in the galaxy than we have so far detected.  The galactic center has also been searched for GWs from unknown neutron stars  \cite{PhysRevD.88.102002}.  Young neutron star systems may be energetic because of the recent supernova and could be strong sources of GW.  The remnant from SN1987A has been searched for GW \cite{PhysRevD.94.082004}.  Finally, the young and active Crab pulsar is observed to be rapidly spinning down.  This is most likely due to electromagnetic radiation.  However, some of this spin down could be due to GW radiation, and LIGO has now limited the spin down power in GW radiation to be less than about 1\% of the total spin down power \cite{2014ApJ...785..119A}.   All of these limits will likely be improved significantly within a few years, because LIGO has recently been upgraded to a more sensitive advanced LIGO configuration.   Indeed, the first advanced LIGO search for GW from known pulsars has now been published \cite{0004-637X-839-1-12}.

We pause now to discuss how the strength of neutron star crust depends on density.   Previous work has focused on the strength of the Coulomb interactions between ions, which increase as the ions move closer together.  Therefore, the strength of the crust is expected to increase with density.  Above a density of about $10^{11}$ g/cm$^3$ the ions become so neutron rich that excess neutrons drip out to form a neutron gas between the ions.  This gas has been ignored in simulations of crust strength \cite{crustbreaking}.  Neutron-ion interactions could possibly change the structure of the body centered cubic (bcc) lattice \cite{Kobyakov2014}.  However MD simulations find little difference in strength between bcc and fcc lattices \cite{crustbreaking}.

Perhaps more interesting is what happens as one approaches nuclear density near $10^{14}$ g/cm$^3$ where the ions start to touch and strongly interact with their neighbors.  Competition between short range nuclear attraction and long range Coulomb repulsion can rearrange nucleons into complex shapes known as nuclear pasta.  This likely greatly changes the material properties.  Nuclear pasta is discussed in the next section.

\section{Nuclear pasta and soft astromaterials}
\label{sec.soft}

An example of a soft astromaterial is ``nuclear pasta," found in the inner crust of neutron stars.

To review, the outer crust of a neutron star is likely a Coulomb crystal, similar in composition to a white dwarf. These Coulomb crystals are made of isolated nuclei and are at densities several orders of magnitude below nuclear saturation. However, the core of a neutron star is believed to be uniform nuclear matter at densities above $3\times 10^{14}$ g/cm$^3$, the nuclear saturation density $n_0$. 
There must therefore be a region in the neutron star crust where nuclear matter transitions from being found in isolated nuclei to existing in bulk.

Under compression, the ions in the crust will rearrange into exotic shapes in order to minimize their energy. The competition between the nuclear attraction of protons and neutrons and the Coulomb repulsion between protons creates a variety of nonspherical nuclei \cite{1984PThPh..71..320H, PhysRevLett.50.2066}.  This transition is now believed to involve several pasta phases.


As a simple illustrative argument, imagine what phases would be seen by an observer descending through the crust, as shown in Fig \ref{fig:NS_PieChart}. At the top of the crust, a thin atmosphere and ocean of ionized nuclei are found embedded in a degenerate Fermi gas of electrons. 
With increasing density these nuclei freeze to form the outer crust.  The outer crust is a solid bcc lattice of nuclei. As the electron Fermi energy increases with depth, electron capture begins to occur creating increasingly neutron rich nuclei. These nuclei eventually reach the neutron drip line and begin to shed neutrons into a free neutron gas surrounding the lattice, marking the transition to the inner crust. 

At the base of the inner crust we find neutron rich nuclei with a free neutron gas. The proton fraction here is near 5\%. At densities above $0.2 n_0$, the nuclei begin to touch and fuse forming complex shapes. As the density approaches $n_0$ the complex shapes transition to uniform nuclear matter.\footnote{Aside from illustrating the phase transitions and anatomy of a neutron star interior, this story of descent also describes the experience of accreted matter. Such material falls onto the neutron star and is buried by later material, thus progressing through the crust until eventually reaching the core. This sort of crust replacement is expected to occur in low mass X-ray binary systems.}
These shapes have since come to be called nuclear pasta, due to their resemblance to spaghetti and lasagna and other namesake pasta. In total, for a 10 km radius NS, the ions of the crust may extend to a depth of about 1 km, and the pasta region may extend an additional 100 m. 

\begin{figure*}[ht!]
\centering
\includegraphics[width=0.95\textwidth]{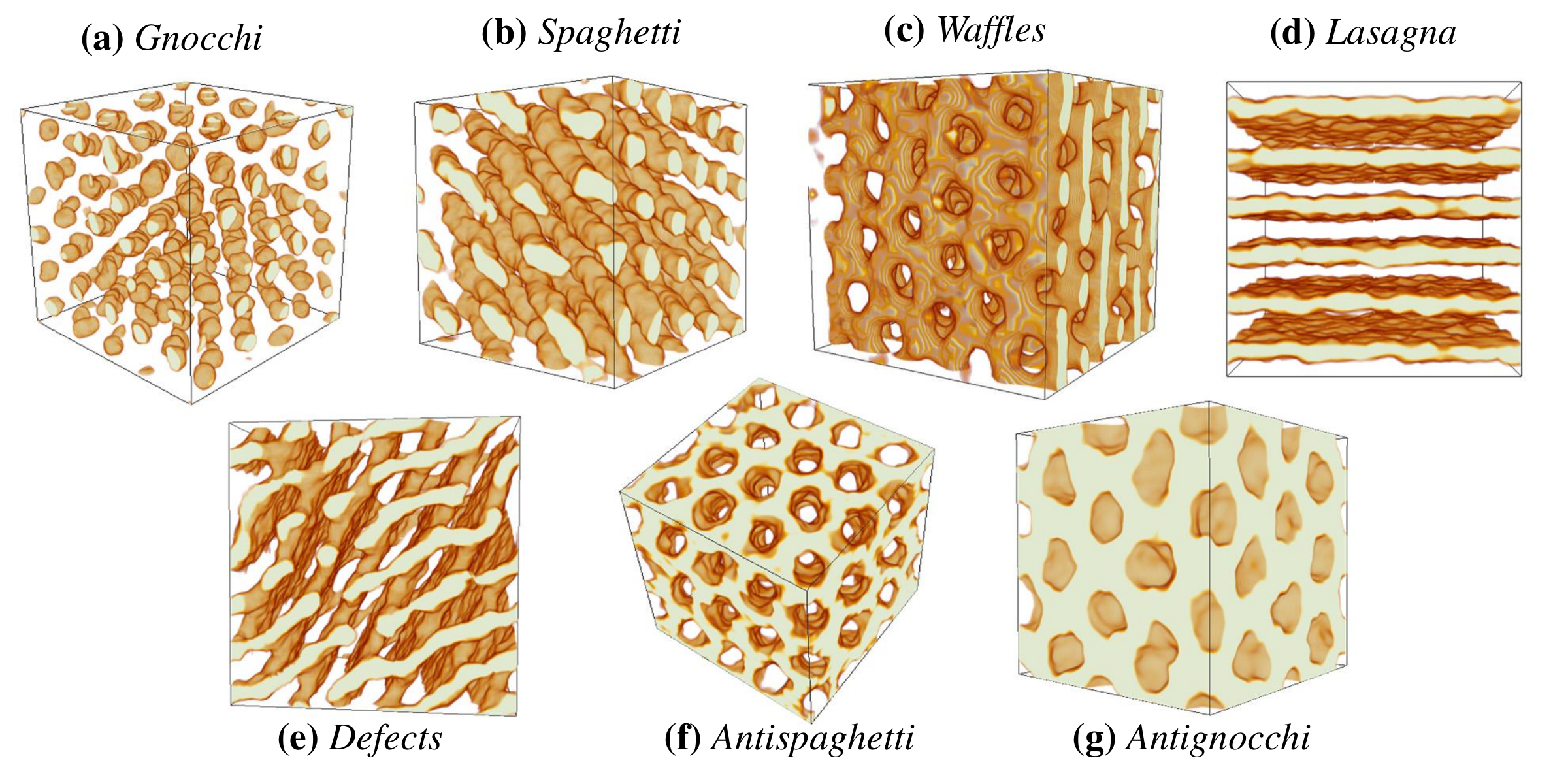}
\caption{\label{fig:7pasta} Nuclear pasta configurations produced in our MD simulations with 51,200 nucleons \cite{PhysRevC.88.065807,PhysRevLett.114.031102,PhysRevC.90.055805}.} 
\end{figure*}

\subsection{Historical Development}

Early work studying nuclear matter at subsaturation density only considered spherical nuclei, nuclear matter with spherical bubbles, and uniform nuclear matter \cite{BAYM1971225}. This work argued that nuclei turned `inside out' at high densities ($0.5 n_0$) to form nuclear matter with spherical bubbles, which is relevant to the equation of state \cite{PhysRevLett.41.1623}.  Later work expanded on this approach using a Skyrme Hamiltonian \cite{LATTIMER1985646}.

The idea of nonspherical nuclei near the saturation density was proposed independently by Ravenhall \etal  (1983) and Hashimoto \etal (1984). This early work used a simple liquid drop model to predict a hierarchy in density of five structures which could be stable in the neutron star crust. These five structures are, in order of increasing density: spheres, cylinders, slabs, cylindrical voids, and spherical voids, and are shown in Fig. \ref{fig:7pasta} (a), (b), (d), (f), and (g). These phases have since come to be called gnocchi, spaghetti, lasagna, antispaghetti, and antignocchi.


Williams and Koonin (1985) used the Thomas-Fermi approximation to a Skyrme functional for symmetric nuclear matter to show that the transitions between these five phases are first-order \cite{WILLIAMS1985844}. Oyamatsu (1993) expanded on this by using Thomas-Fermi-Skyrme to study neutron star matter in beta equilibrium, confirming the stability of the standard hierarchy in neutron star matter \cite{OYAMATSU1993431}. Lastly, Lorenz \etal confirmed these results using Hartree-Fock \cite{PhysRevLett.70.379}.

Time-dependent Hartree-Fock has recently been used to simulate small numbers of nucleons (up to 2,000) which form pasta and have been used to produce phase diagrams of nuclear pasta for a variety of temperatures, densities, and proton fractions \cite{Magierski2002,PhysRevC.76.024312,PhysRevC.79.055801,PhysRevC.77.035806, PhysRevC.87.055805}. These simulations have been useful for showing the existence of pasta phases beyond the standard hierarchy, such as the networked gyroid structure \cite{PhysRevC.91.025801, PhysRevC.92.045806}.


Additionally, semi-classical molecular dynamics (MD) models have been developed, originally for heavy ion collisions \cite{AICHELIN198614,PhysRevC.31.1783,DORSO1987287}, and then applied to nuclear pasta \cite{PhysRevC.57.655}.  Watanabe, Sonoda, and their collaborators have done extensive work using MD to determine the structure and possible formation of nuclear pasta \cite{PhysRevC.66.012801,PhysRevC.68.035806,PhysRevC.69.055805,PhysRevLett.94.031101,PhysRevLett.103.121101}.   We developed an MD model, originally to study the neutrino opacity of nuclear pasta, although the model has since been used to calculate many other pasta properties \cite{Horowitz:2004yf,PhysRevC.69.045804}.


Despite being (semi-)classical these MD simulations have been able to robustly reproduce the predicted phases of nuclear pasta, see for example \cite{PhysRevLett.114.031102, PhysRevC.91.065802}. Additionally, classical simulations require considerably less computation time than quantum simulations, so they can be run for much larger simulation volumes. This has enabled the identification of a variety of new pasta phases such as waffles \cite{PhysRevC.90.055805} and defects \cite{PhysRevLett.114.031102}, which are shown in Fig. \ref{fig:7pasta} (c) and (e). In addition, the classical simulations have also been used to generate initial conditions for the quantum simulations, which reduce the computation time considerably \cite{PhysRevC.93.055801}.

\subsection{Molecular Dynamics}

A large body of recent work has used semi-classical molecular dynamics simulations to study nuclear pasta structures. We illustrate this work by describing a semi-classical model   as one example \cite{Horowitz:2004yf}.  This model treats protons and neutrons as point particles, and uses a set of three classical two-body potentials for describing their interaction:
\begin{subequations}
\begin{align}
 V_{np}(r)&=a e^{-r^2/\Lambda}+[b-c]e^{-r^2/2\Lambda}\,  ,\\
 V_{nn}(r)&=a e^{-r^2/\Lambda}+[b+c]e^{-r^2/2\Lambda}\, ,\\
 V_{pp}(r)&=a e^{-r^2/\Lambda}+[b+c]e^{-r^2/2\Lambda}+\frac{\alpha}{r}e^{-r/\lambda}\, ,
\end{align}
\end{subequations}
where the subscripts $n$ and $p$ denote the interactions between neutrons and protons and $r$ is the inter-particle separation. The nucleons interact by a short range potential meant to model the nuclear interaction whose strength and range are determined by the parameters $a$, $b$, $c$ and $\Lambda$, which are given in Table \ref{Tab:parameters}. These parameters were chosen to approximately reproduce the saturation density and binding energy per nucleon of nuclear matter, the energy of neutron matter at saturation density and the binding energies of a few select nuclei \cite{PhysRevC.69.045804}.  They have been found to reproduce nuclear statistical equilibrium for simulations at low densities \cite{PhysRevC.91.065802}.  The nuclear potentials have an intermediate range attraction, a short range repulsion, and the protons interact with an additional long ranged Coulomb repulsion. 

The use of a semi-classical approximation deserves comment.  Individual nucleons are light and have important quantum zero point motions.  However, we are most interested in the large scale behavior that typically involves large clusters involving thousands of nucleons.  These clusters are heavy and therefore behave classically.  The most important quantum effects, at small scales, can be mocked up by choosing values for the parameters $a$, $b$, $c$, and $\Lambda$ as described above. 
Electrons are not treated explicitly, and are instead included by adding a screening factor to the proton-proton Coulomb interaction. This screening factor is taken to be the Thomas-Fermi screening length $
\lambda$, see Eq. \ref{eq.lambda}.

\begin{table}[h]
\caption{\label{Tab:parameters} Parameters of the nuclear interaction. The strength of the short-range repulsion between nucleons is given by $a$, while $b$ and $c$ set the strength of the intermediate-range attraction. The characteristic length scale of the nuclear potential is given by  $\Lambda$ .}
\begin{ruledtabular}
\begin{tabular}{*{4}{c}}
$a$ (MeV) &$b$ (MeV)&$c$ (MeV)&$\Lambda$ (fm$^{2}$) \\
\hline
  110     &  $-$26    &   24    &    1.25      \\
\end{tabular}
\end{ruledtabular}
\end{table}


These potentials were first developed by \cite{Horowitz:2004yf} in order to study how pasta affects neutrino transport in supernovae. In that work, Horowitz \etal calculated the static structure factor $S_n(q)$, allowing them to determine the mean free path of neutrinos, see Sec. \ref{ss:Sq}.   MD simulations can be performed with many nucleons, and simulations with up to 409,600 nucleons have been reported in the literature to study finite size effects. 

\subsubsection{Topology}

The pasta phases and their transitions can be rigorously quantified by their geometry and topology. The Minkowski functionals offer a powerful tool to describe the morphology of pasta structures \cite{PhysRevC.66.012801,PhysRevC.77.035806,PhysRevC.87.055805}. There are 4 Minkowski functionals defined in 3 dimensions; these are the volume $V$, surface area $A$, mean breadth $B$, and Euler characteristic $\chi$.

The volume is obtained straightforwardly as the occupied fraction of the simulation. The remaining three require isosurfaces to be generated for constant density so that we can study the distribution of protons. For our simulations we use a threshold proton density of $n=0.030 \text{ fm}^{-3}$ protons \cite{PhysRevC.88.065807}.  Note that there is often more contrast in the proton density then in the neutron density. 

The Minkowski functionals for these surfaces are calculated by integrating their principle curvatures, $k_1$ and $k_2$, over the surface area, as shown in Tab. \ref{Tab:minkowski}. Recall that positive curvature is convex, while negative curvature is concave. The sign of two Minkowski functionals, $B$ and $\chi$, is often sufficient to characterize the phase of the pasta. 

\begin{table}[h]
\caption{\label{Tab:minkowski} Minkowski functionals in 3 dimensions. $K$ is the domain over which the principle curvatures, $k_1$ and $k_2$, are evaluated. }
\begin{ruledtabular}
\begin{tabular}{*{2}{l}}
$V$ 	& Volume  \\
$A = \int_{\partial K} dA $  	&   Surface Area \\
$B = \int_{\partial K} (k_1 + k_2 )/4 \pi  dA $  &	 Mean Breadth       \\
$\chi = \int_{\partial K} (k_1 \cdot k_2 )/4 \pi dA $   &	  Euler Characteristic      \\
\end{tabular}
\end{ruledtabular}
\end{table}


The mean breadth can be thought of as a measure of the convexity or concavity of the pasta. At high density we expect the pasta will contain voids, which are concave, while at low density the pasta will contain many isolated structures (nuclei) which are convex. 

The Euler characteristic is unique among the Minkowski functionals in that it quantifies the topology.  For a periodic surface, the Euler characteristic is proportional to the number of connected components plus the number of cavities minus the number of tunnels (through the system)
\begin{equation}
\begin{split}
 \chi =  \text{ \#(Connected Components) } \\ +  \text{ \#(Cavities) } - \text{ \#(Tunnels)}.
\end{split}
\end{equation}
Taken together, the mean breadth and Euler characteristic are sufficient to quantify the phases of nuclear pasta. Additionally, when the Minkowski functionals are plotted as a function of density, discontinuities in the curvature are observed which correspond to phase transitions in the pasta. By treating curvature as a thermodynamic variable in this way we are able to quantify the phase transitions in nuclear pasta and construct phase diagrams for our model.

\begin{table}[]
\centering
\caption{The eight topological classifications of periodic structures based on the sign of the Euler characteristic $\chi$ and the mean breadth B.}
\label{tab:topology}
\begin{ruledtabular}
\begin{tabular}{c|ccc}
            & $ B < 0$ & B $\sim$ 0   & $B > 0$  \\
\hline
$\chi > 0 $         & sph b      &       & sph            \\
$\chi \sim 0 $   & rod-1 b    & slab  & rod-1          \\
$\chi < 0 $         & rod-2 b    & rod-3 & rod-2        
\end{tabular}
\end{ruledtabular}

\end{table}

\begin{figure}[ht!]
\centering
\includegraphics[width=0.5\textwidth]{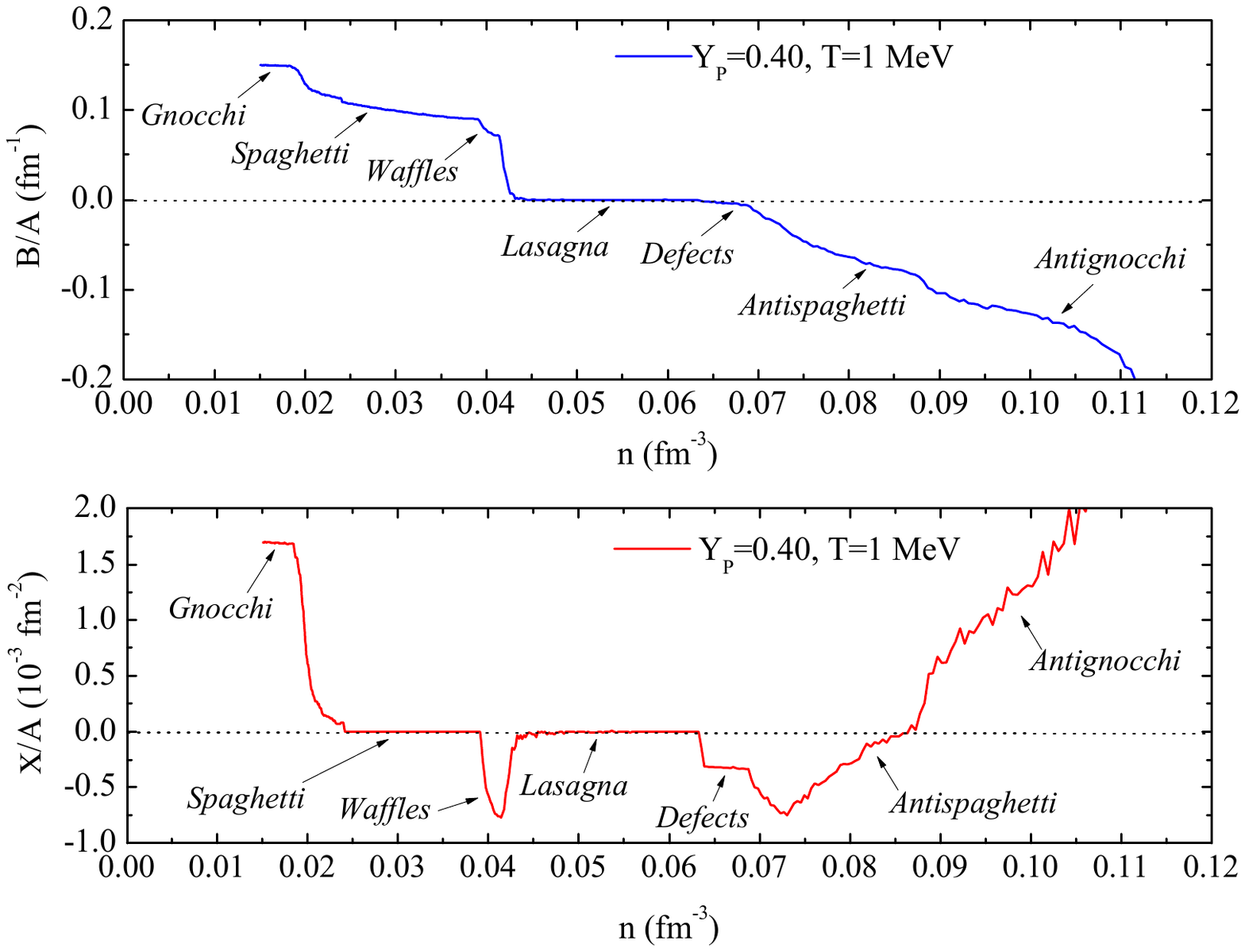}
\caption{\label{fig:minkowski} The normalized Minkowski functionals $B/A$ and $X/A$ as a function of density for a simulation of 51,200 nucleons.} 
\end{figure}

In Fig. \ref{fig:minkowski} we show the Minkowski functionals as calculated for one simulation at $Y_P = 0.4$ and $T=1$ MeV with 51,200 nucleons. This simulation was initialized at a density 0.12 fm$^{-3}$ and the simulation volume was expanded very slowly at a constant rate over $5\times 10^7$\ MD timesteps until it reached a density of 0.015 fm$^{-3}$.

There is excellent agreement between the functionals in indicating phase transitions. While one Minkowski functional is sufficient to show most of the phase transitions, when taken together the interpretation becomes more clear. For example, the mean breadth does not change sharply between densities of 0.07 and 0.065 fm$^{-3}$, but the Euler characteristic shows a clear shift in topology, indicating the presence of `defects' connecting the lasagna plates. Similarly, the mean breadth only weakly shows a transition near densities of 0.042 fm$^{-3}$ but a sharp decline in the Euler characteristic indicates that holes have formed in the lasagna, which we call the waffle phase. The transitions in the Minkowski functionals are also better resolved for slower expansion rates. 

Ultimately, it is useful to discriminate between pasta phases so that we can study geometric dependencies of their transport properties using the static structure factor. 

\subsubsection{Static Structure Factor}
\label{ss:Sq}

The static structure factors $S_p({\bf q})$ and $S_n({\bf q})$ are used to describe how leptons scatter from nucleons in the pasta, which affects observable properties of neutron stars, as described in Sec. \ref{ss:observables}. In short, neutrinos scatter primarily from neutrons which affects supernova neutrino transport, and electrons scatter from protons which affects the thermal conductivity, electrical conductivity, and shear viscosity of the crust.

The proton static structure factor $S_p({\bf q})$ is a coherent sum of amplitudes for scattering from all of the protons in the medium, and is defined as 
\begin{equation}\label{eq:sq}
    S_p(\mathbf{q}) = \frac{1}{N} \left \langle \sum_{j,k=1}^N \mathrm{e}^{-i \mathbf{q}\cdot (\mathbf{r}_j - \mathbf{r}_k)} \right \rangle
\end{equation}
where $j$ and $k$ are summed over protons, $r_j$ and $r_k$ are proton positions, ${\bf q}$ is the momentum transfer, and $N$ is the number of protons in the simulation. The brackets indicate a time average during an MD simulation.  The neutron static structure factor $S_n({\bf q})$ is defined in a similar way as a sum over neutrons.

\begin{figure}[ht!]
\centering
\includegraphics[width=0.5\textwidth]{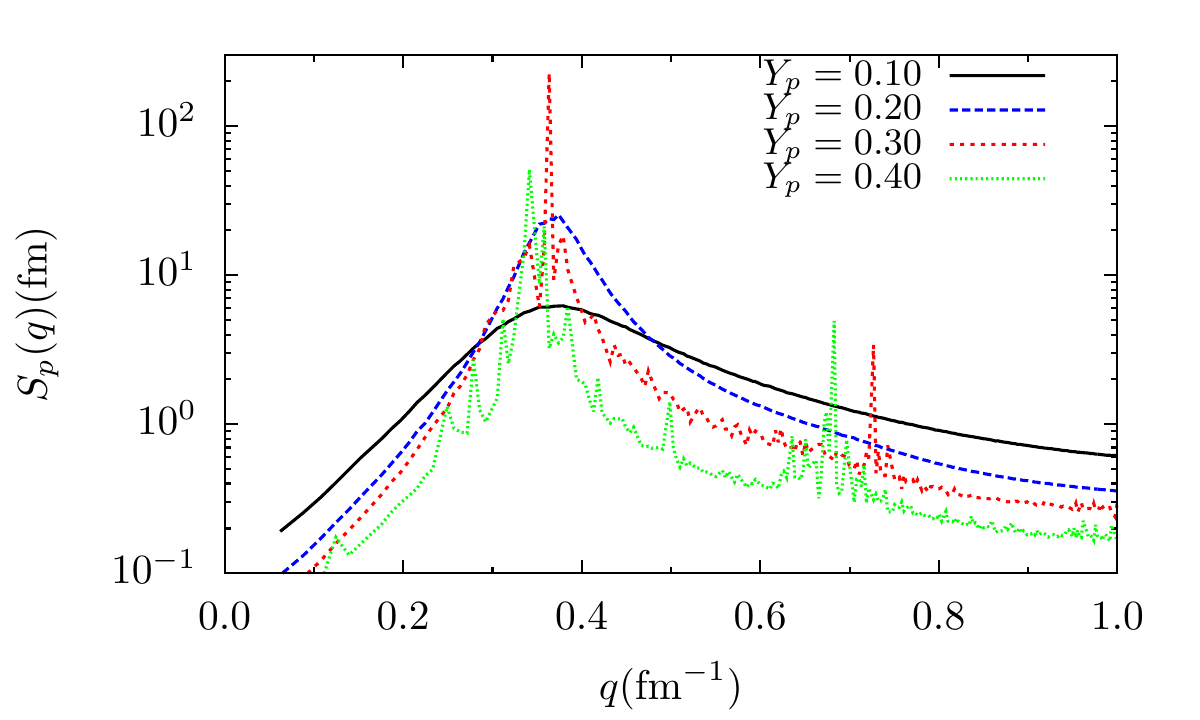}
\caption{\label{fig:sq} The angle-averaged proton static structure factor $S_p(q)$ versus momentum transfer $q$.  This is for nuclear pasta with $n = 0.050$ fm$^{-3}$, temperature $T=1$ MeV, and proton fractions $Y_P$ = 0.10, 0.20, 0.30, and 0.40. Reproduced from \cite{PhysRevC.90.055805}.} 
\end{figure}

An example of angle-averaged $S_p(q)$ for protons can be seen in Fig. \ref{fig:sq}.  At this density of $n=0.05$ fm$^{-3}$, pasta is in the waffle phase for a proton fraction $Y_p=0.3$ and in the lasagna phase for $Y_P$=0.40.   The peaks in $S_p(q)$ near $q=0.35$ fm$^{-1}$ and 0.70 fm$^{-1}$ correspond to Bragg scattering from the pasta planes. 

Once the proton structure factor has been obtained it may be used to calculate the shear viscosity $\eta$, electrical conductivity $\sigma$ and thermal conductivity $\kappa$ by integrating $S_p(q)$ over $q$ \cite{PhysRevC.78.035806}.   For example the thermal conductivity $\kappa$ is given by,
\begin{equation}
\kappa\approx \frac{\pi k_F k^2 T}{12 \alpha^2 \Lambda_\kappa}\, ,
\end{equation}
where the Coulomb logarithm $\Lambda_\kappa$ describes electron-pasta scattering \cite{PhysRevC.78.035806},
\begin{equation}
\Lambda_\kappa\approx \int^{2k_F}_0\frac{dq}{q}(1-\frac{q^2}{4k_F^2})S_p(q)\, .
\end{equation}
The neutrino opacity of the pasta can similarly be calculating by integrating the neutrino-single nucleon scattering cross section, weighted by $S_n(q)$ over $q$ \cite{PhysRevC.69.045804}. Once calculated, these parameters allow us to interpret observable properties of neutron stars. 

\subsection{Observables}\label{ss:observables}

In this section we discuss astrophysical implications of the presence of a pasta layer. Specifically, we discuss the relevance of pasta to the scattering of supernova neutrinos in Sec. \ref{ss:neutrinopasta},  pulsar spin periods and magnetic field decay  in Sec. \ref{ss:spinperiod}, and to late time crust cooling in Sec. \ref{ss:crustcooling}. 

\subsubsection{Supernova neutrino-pasta scattering}
\label{ss:neutrinopasta}
The structure of conventional materials can be studied with X-ray diffraction.  For example, the Advanced Photon Source (APS) at Argonne National Laboratory produces well collimated beams of perhaps $10^{15}$ X-ray photons per second.  

Similarly, it may be possible to study astromaterials with neutrino scattering.   A core collapse supernova (SN), the gigantic explosion of a massive star, provides an extraordinarily intense source of $10^{57}$ neutrinos per second.   The coherent scattering of these neutrinos (from the complex shapes) 
may provide evidence of nuclear pasta.  Indeed neutrino-pasta scattering may slow diffusion and greatly increase the neutrino signal at late times of 10 or more seconds after stellar core collapse \cite{Horowitz:2016fpa}.  Supernova neutrinos are observed in large underground detectors such as Super-Kamiokande \cite{Scholberg2012}.  Thirty years ago, we saw about 20 events from SN1987a \cite{PhysRevD.70.043006}.  We expect many thousands of events from the next galactic core collapse SN, which will allow us to probe SN and proto-NS interiors \cite{Scholberg2012}.

\subsubsection{Spin Period}\label{ss:spinperiod}

The existence of nuclear pasta in the inner crust has been proposed to explain the absence of isolated slowly spinning ($P>12 s$) X-ray pulsars \cite{2013NatPh...9..431P}. 

Young neutron stars with strong magnetic fields ($B > 10^{14} G$), called magnetars, are observed to be have spin periods $P$ of 2-12 s. These neutron stars spin-down by emitting magnetic dipole radiation, thus losing rotational energy as electromagnetic radiation. 

Given the strength of their magnetic fields one would expect that their rotational period would decrease quickly after formation, reaching rotational periods of 20-30 s. However, no such population of isolated slowly rotating neutron stars is observed. 

This can be reconciled by supposing the magnetic field decays quickly, in approximately 1 Myr. For this to be the case, there must be an electrically resistive component to the neutron star that causes the electric currents supporting the magnetic field to dissipate.

This resistive component cannot be found in the core; the protons are superconducting and thus there will be little dissipation of the currents. If the crust is resistive then electric currents can decay by losing energy to Joule heating of the crust. In this case, electron scattering from impurities is the dominant dissipative process. 

At the base of the inner crust near the crust-core boundary, we find the pasta layer which can be heterogeneous. Nuclear matter in the pasta layer is expected to have complex shapes with unusual charge distributions. Even symmetric pasta phases may have defects, effectively acting like impurities for electron scattering \cite{PhysRevC.93.065806}. 

Pons considers magnetic field configurations in the crust that don't penetrate into the core. Then, assuming that the pasta layer has a high impurity parameter compared to the rest of the star, Pons shows that the magnetic field will decay by one to two orders of magnitude within 0.1-1 Myr. This would produce a population of isolated X-ray pulsars all with spin periods less than 12 s, consistent with observations. Pons' assumption that the pasta has a high impurity parameter is consistent with the results of our molecular dynamics simulations, which suggest that pasta may be well approximated by an impurity parameter of 30-40 \cite{PhysRevLett.114.031102,PhysRevC.93.065806}

\subsubsection{Crust Cooling}\label{ss:crustcooling}

As we saw in Sec. \ref{ss:spinperiod}, an impure pasta layer with a low electrical conductivity can dissipate electric currents, causing magnetic fields to decay. In this section, we discuss how the low thermal conductivity of pasta can affect cooling in quiescent low mass X-ray binaries (LMXB) - these are neutron stars with a main sequence companion. 

For conventional metals at constant temperature the electrical conductivity is linearly related to the thermal conductivity as described by the Wiedemann-Franz law. This is because free electrons in the metal are both a thermal and electrical carrier. Similarly, in the neutron star, electrons can serve as both a thermal and electrical carrier in the crust. Thus, we consider an analogous approach to Pons \etal to study crust cooling; what if the pasta is thermally resistive?

In an accreting LMXB mass falls from a companion onto a neutron star and emits X-rays. If the accretion stops then the neutron star will cool, and the time evolution of the surface temperature can be used to probe the thermal conductivity deep in the crust. 

Observations have been made of MXB 1659-29 for ten years into quiescence. The surface temperature 3-10 years into quiescence is sensitive to the thermal conductivity at densities and depths where we expect to find nuclear pasta, because it takes a cooling wave this long to diffuse from the surface \cite{PhysRevLett.114.031102}.  

\begin{figure}[ht!]
\centering
\includegraphics[width=0.45\textwidth]{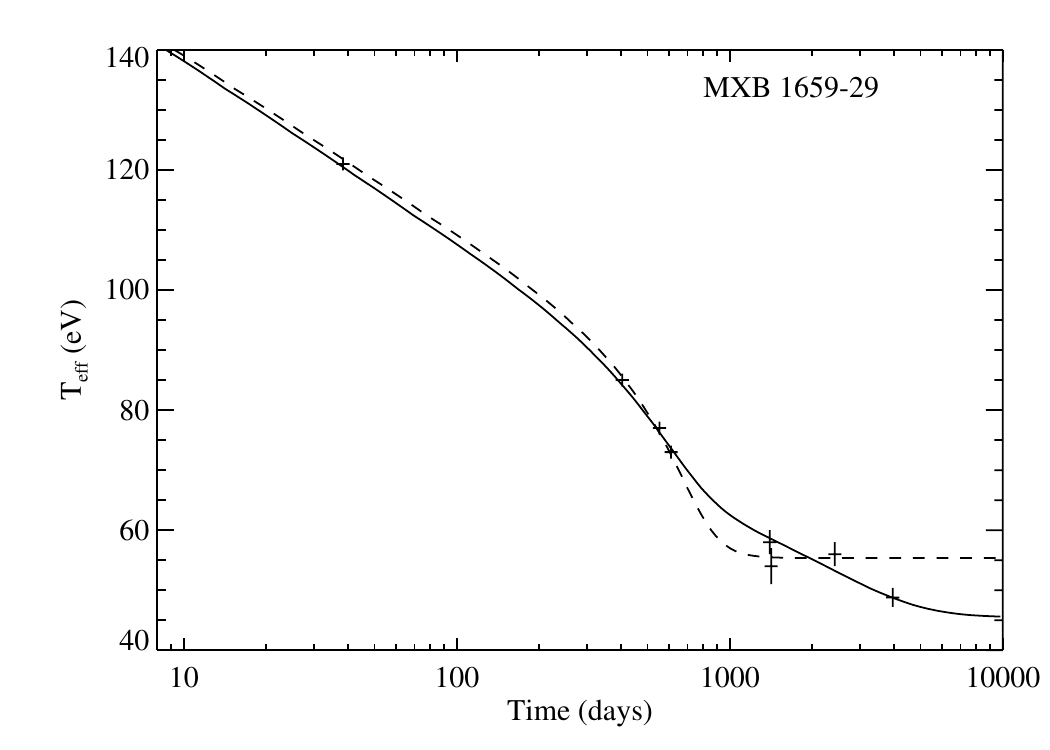}
\caption{\label{fig:1659cooling} Surface temperature of MXB 1659-29 vs time since accretion stopped.  The dashed line shows the predicted cooling curve for a model without a resistive pasta layer while the solid line shows the predicted cooling curve for a model with a resistive pasta layer. (Reproduced from \cite{PhysRevLett.114.031102})} 
\end{figure}

In Fig. \ref{fig:1659cooling} we show the evolution of the observed surface temperature of MXB 1659-29 for ten years into quiescence with cooling curves for two different models. The dashed curve shows the standard cooling model without a thermally resistive layer at the base of the crust. In this model, the crust should have come into equilibrium with the core approximately 1000 days into quiescence, which does not fit the most recent observation which shows that the surface temperature has continued to drop. This observation of late time cooling implies the presence of a resistive layer deep the crust, which is consistent with our interpretation of pasta as a thermal resistor.

\subsection{Relation to Soft Condensed Matter}

There is an interesting analogy between nuclear pasta and soft matter physics, a broad field which studies the properties of a variety of materials such as glasses, liquid crystals, and biological materials.   Pethick and Ravenhall discuss several condensed matter properties of neutron star crust matter \cite{pethick1995}, while Pethick and Potekhin consider nuclear pasta as a liquid crystal \cite{Pethick19987}.  Watanabe and Sonoda discuss nuclear pasta as soft condensed matter  \cite{Watanabe:2005qt}, see also \cite{WATANABE2000455}.  Possible gyroid phases of nuclear pasta are considered in \cite{PhysRevLett.103.132501,PhysRevC.91.025801}.  These are periodic networklike structures with negatively curved surfaces. 

At low temperature, approximately 0.5 MeV, our nuclear pasta model is known to freeze and crystallize \cite{PhysRevC.91.065802}. Other models of nuclear pasta exhibit this behavior as well, and have been shown to exhibit a first order phase transition at low temperature \cite{PhysRevC.90.065803}. Our nuclear pasta model has been found to have a second order phase transition for fast cooling rates and high proton fraction, similar to the quenching of a liquid to a glass. This is not surprising; the functional form of our nuclear pasta model is comparable to a binary Lennard-Jones interaction with a Coulomb interaction, such as the Kob-Andersen mixture which is known to undergo a glass transition \cite{PhysRevLett.73.1376}. 

In our work, this glass transition is an artifact of the classical model and is not believed to be present in neutron stars, so we limit our study of pasta to temperatures above this transition. However, glasses are a major topic in soft matter physics, and our model (with appropriately scaled units) could be useful to those seeking to study them.

The soft matter analogue can also be seen at higher temperatures from the self assembly of nuclear pasta. 

In living systems, phospholipids are assembled into cellular organelle membranes with a variety of geometries. These phospholipids have a hydrophilic head with two hydrophobic tails. In an aqueous solution, the heads are attracted to the water while the tails are repelled, causing them to aggregate and form vesicles which bound a volume. The relative abundance of water and phospholipids determines the resulting shapes of these vesicles, which can be spheres, cylinders, plates, and matter with voids, similar to nuclear pasta \cite{SEDDON19901,buehler2015cell}. 

In both systems the emergence of these geometries has a common origin in \textit{frustration}. In the pasta model, competition between the nuclear attraction and Coulomb repulsion must be minimized. In the phospholipids the interaction between the hydrophilic heads and hydrophobic tails with the water, and the bending energy of the membrane, must be minimized. It is remarkable that despite the differences in the interactions at play, and a difference in density of $10^{14}$, both systems have this coarse grained self assembly in common. 

\begin{figure}[ht!]
\centering
\includegraphics[width=0.45\textwidth]{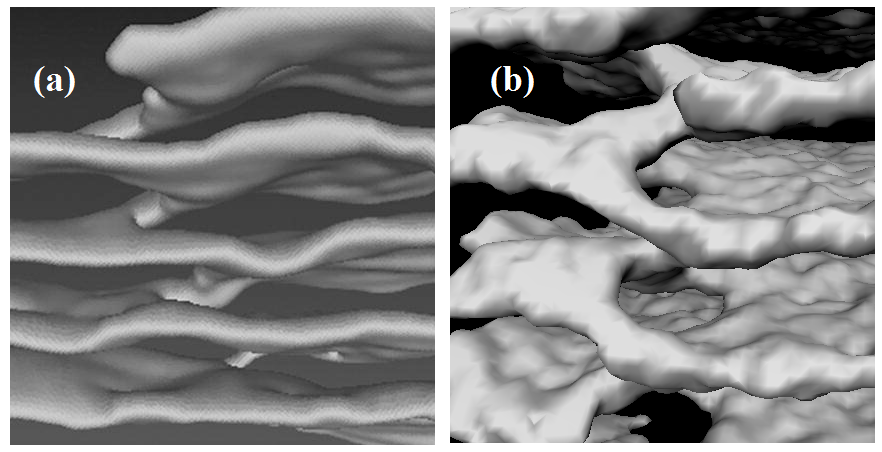}
\caption{\label{fig:ramps} Comparison of helicoidal ramps in (a) electron micrograph of endoplasmic reticulum at a density near 1 g/cm$^3$ (reproduced from \cite{Terasaki2013285}) and (b) nuclear pasta simulations at a density of $10^{14}$ g/cm$^3$. } 
\end{figure}

As one example, the `defects' that we observe in our simulations have a distinct chirality and they spontaneously organize to form dipoles, quadrupoles, and octopoles of alternating left and right handedness. Recently, a geometrically comparable structure has been observed in the endoplasmic reticulum of mouse cells, which can be seen in Fig. \ref{fig:ramps} \cite{Terasaki2013285}. In both systems the parallel plates are connected by helicoidal ramps which are geometrically similar to spiral ramps in parking garages \cite{Horowitz:2015gda}. 
These helicoids are found in dipoles or quadrupoles of alternating helicity. Additionally, the dipolar helicoids are found at an approximately 45 degree angle with respect to the parallel plates, and the quadrupoles are oriented at 90 degrees with respect to the plates. 

We see from this correspondence that the self assembly physics seen in our pasta model belongs to a more general class of materials, such as gels and liquid crystals, and perhaps our model (with appropriately scaled units) can be useful to soft matter physicists seeking to simulate these systems. 

\section{Summary and perspective}
\label{conclusions}
For this colloquium, we defined `astromaterial science' as the study of materials, in astronomical objects, that are qualitatively denser than materials on earth.  We discussed one hard astromaterial, Coulomb crystals, and one soft astromaterial, nuclear pasta. 

Coulomb crystals are present in the cores of white dwarfs and the crusts of neutron stars, forming when the stars cool and freeze. Our large-scale molecular dynamics simulations of the breaking stress of neutron star (NS) crust suggest it is the strongest material known, some ten billion times stronger than steel.  This strong crust can support large mountains that, on a rapidly rotating neutron star, can be a source of detectable gravitational waves.  

Nuclear pasta is an example of a soft astromaterial. It is expected near the base of the NS crust. Competition between nuclear attraction and Coulomb repulsion rearranges neutrons and protons into complex non-spherical shapes such as flat sheets (lasagna) or thin tubes (spaghetti).  Our semi-classical molecular dynamics simulations of nuclear pasta can be used to calculate transport properties including shear viscosity, thermal conductivity, electrical conductivity, and neutrino opacity, which are relevant for interpreting a variety of observations of neutron stars, such as spin-down and cooling of accreting binaries in quiescence. 

In the near future, one should calculate additional nuclear pasta properties such as the bulk viscosity, shear modulus, and breaking strain.  The shear modulus is important for NS crust oscillation frequencies, while the bulk viscosity could dampen collective oscillations.  Finally, the breaking strain is important for crust mountains and for starquakes.

New multi-messenger observations will advance astromaterial science.  The NICER X-ray telescope is scheduled to be installed soon on the International Space Station \cite{NICER}.  NICER aims to measure NS radii and constrain the equation of state (pressure versus density) of neutron rich dense matter.  Additional X-ray observations of transiently accreting NS will improve our knowledge of crust properties including the thermal conductivity.  Gravitational wave (GW) observations of binary NS mergers can constrain the radius and polarizability of NS.  Searches for continuous GW from rotating NS will constrain NS shapes and provide strict limits on crust mountains.  Finally, we expect many thousands of neutrino events from the next galactic core collapse Supernova (SN).  This could provide strong evidence of nucleons clustering into large nuclear pasta shapes.

\section{Acknowledgments}

We thank Greg Huber for useful discussions.  This research was supported in part by DOE grants DE-FG02-87ER40365 (Indiana University) and DE-SC0008808 (NUCLEI SciDAC Collaboration). 
Computer time was provided by the INCITE program. This research used resources of the Oak Ridge Leadership Computing Facility located at Oak Ridge National Laboratory, which is supported by the Office of Science of the Department of Energy under Contract No. DEAC05-00OR22725.
This research was supported in part by Lilly Endowment, Inc., through its support for the Indiana University Pervasive Technology Institute, and in part by the Indiana METACyt Initiative. The Indiana METACyt Initiative at IU is also supported in part by Lilly Endowment, Inc.       This material is based upon work supported by the National Science Foundation under Grant No. CNS-0521433


\begin{thebibliography}{102}%
\makeatletter
\providecommand \@ifxundefined [1]{%
 \@ifx{#1\undefined}
}%
\providecommand \@ifnum [1]{%
 \ifnum #1\expandafter \@firstoftwo
 \else \expandafter \@secondoftwo
 \fi
}%
\providecommand \@ifx [1]{%
 \ifx #1\expandafter \@firstoftwo
 \else \expandafter \@secondoftwo
 \fi
}%
\providecommand \natexlab [1]{#1}%
\providecommand \enquote  [1]{``#1''}%
\providecommand \bibnamefont  [1]{#1}%
\providecommand \bibfnamefont [1]{#1}%
\providecommand \citenamefont [1]{#1}%
\providecommand \href@noop [0]{\@secondoftwo}%
\providecommand \href [0]{\begingroup \@sanitize@url \@href}%
\providecommand \@href[1]{\@@startlink{#1}\@@href}%
\providecommand \@@href[1]{\endgroup#1\@@endlink}%
\providecommand \@sanitize@url [0]{\catcode `\\12\catcode `\$12\catcode
  `\&12\catcode `\#12\catcode `\^12\catcode `\_12\catcode `\%12\relax}%
\providecommand \@@startlink[1]{}%
\providecommand \@@endlink[0]{}%
\providecommand \url  [0]{\begingroup\@sanitize@url \@url }%
\providecommand \@url [1]{\endgroup\@href {#1}{\urlprefix }}%
\providecommand \urlprefix  [0]{URL }%
\providecommand \Eprint [0]{\href }%
\providecommand \doibase [0]{http://dx.doi.org/}%
\providecommand \selectlanguage [0]{\@gobble}%
\providecommand \bibinfo  [0]{\@secondoftwo}%
\providecommand \bibfield  [0]{\@secondoftwo}%
\providecommand \translation [1]{[#1]}%
\providecommand \BibitemOpen [0]{}%
\providecommand \bibitemStop [0]{}%
\providecommand \bibitemNoStop [0]{.\EOS\space}%
\providecommand \EOS [0]{\spacefactor3000\relax}%
\providecommand \BibitemShut  [1]{\csname bibitem#1\endcsname}%
\let\auto@bib@innerbib\@empty
\bibitem [{\citenamefont {{Aasi}}\ \emph {et~al.}(2014)\citenamefont {{Aasi}},
  \citenamefont {{Abadie}}, \citenamefont {{Abbott}}, \citenamefont {{Abbott}},
  \citenamefont {{Abbott}}, \citenamefont {{Abernathy}}, \citenamefont
  {{Accadia}}, \citenamefont {{Acernese}}, \citenamefont {{Adams}},
  \citenamefont {{Adams}},\ and\ \citenamefont {et~al.}}]{2014ApJ...785..119A}%
  \BibitemOpen
  \bibfield  {author} {\bibinfo {author} {\bibnamefont {{Aasi}}, \bibfnamefont
  {J}}, \bibinfo {author} {\bibfnamefont {J.}~\bibnamefont {{Abadie}}},
  \bibinfo {author} {\bibfnamefont {B.~P.}\ \bibnamefont {{Abbott}}}, \bibinfo
  {author} {\bibfnamefont {R.}~\bibnamefont {{Abbott}}}, \bibinfo {author}
  {\bibfnamefont {T.}~\bibnamefont {{Abbott}}}, \bibinfo {author}
  {\bibfnamefont {M.~R.}\ \bibnamefont {{Abernathy}}}, \bibinfo {author}
  {\bibfnamefont {T.}~\bibnamefont {{Accadia}}}, \bibinfo {author}
  {\bibfnamefont {F.}~\bibnamefont {{Acernese}}}, \bibinfo {author}
  {\bibfnamefont {C.}~\bibnamefont {{Adams}}}, \bibinfo {author} {\bibfnamefont
  {T.}~\bibnamefont {{Adams}}}, \ and\ \bibinfo {author} {\bibnamefont
  {et~al.}}} (\bibinfo {year} {2014}),\ \bibfield  {title} {\enquote {\bibinfo
  {title} {{Gravitational Waves from Known Pulsars: Results from the Initial
  Detector Era}},}\ }\href {\doibase 10.1088/0004-637X/785/2/119} {\bibfield
  {journal} {\bibinfo  {journal} {\apj}\ }\textbf {\bibinfo {volume} {785}},\
  \bibinfo {eid} {119}},\ \Eprint {http://arxiv.org/abs/1309.4027}
  {arXiv:1309.4027 [astro-ph.HE]} \BibitemShut {NoStop}%
\bibitem [{\citenamefont {Aasi}\ and\ \citenamefont
  {et~al.}(2013)}]{PhysRevD.88.102002}%
  \BibitemOpen
  \bibfield  {author} {\bibinfo {author} {\bibnamefont {Aasi}, \bibfnamefont
  {J}}, \ and\ \bibinfo {author} {\bibnamefont {et~al.}} (\bibinfo
  {collaboration} {LIGO Scientific Collaboration and Virgo Collaboration})}
  (\bibinfo {year} {2013}),\ \bibfield  {title} {\enquote {\bibinfo {title}
  {Directed search for continuous gravitational waves from the galactic
  center},}\ }\href {\doibase 10.1103/PhysRevD.88.102002} {\bibfield  {journal}
  {\bibinfo  {journal} {Phys. Rev. D}\ }\textbf {\bibinfo {volume} {88}},\
  \bibinfo {pages} {102002}}\BibitemShut {NoStop}%
\bibitem [{\citenamefont {Abbott}\ \emph
  {et~al.}(2017{\natexlab{a}})\citenamefont {Abbott}, \citenamefont {Abbott},
  \citenamefont {Abbott}, \citenamefont {Abernathy}, \citenamefont {Acernese},
  \citenamefont {Ackley}, \citenamefont {Adams}, \citenamefont {Adams},
  \citenamefont {Addesso}, \citenamefont {Adhikari}, \citenamefont {Adya},
  \citenamefont {Affeldt}, \citenamefont {Agathos}, \citenamefont {Agatsuma},
  \citenamefont {Aggarwal}, \citenamefont {Aguiar},\ and\ \citenamefont
  {Aiello}}]{0004-637X-839-1-12}%
  \BibitemOpen
  \bibfield  {author} {\bibinfo {author} {\bibnamefont {Abbott}, \bibfnamefont
  {B~P}}, \bibinfo {author} {\bibfnamefont {R.}~\bibnamefont {Abbott}},
  \bibinfo {author} {\bibfnamefont {T.~D.}\ \bibnamefont {Abbott}}, \bibinfo
  {author} {\bibfnamefont {M.~R.}\ \bibnamefont {Abernathy}}, \bibinfo {author}
  {\bibfnamefont {F.}~\bibnamefont {Acernese}}, \bibinfo {author}
  {\bibfnamefont {K.}~\bibnamefont {Ackley}}, \bibinfo {author} {\bibfnamefont
  {C.}~\bibnamefont {Adams}}, \bibinfo {author} {\bibfnamefont
  {T.}~\bibnamefont {Adams}}, \bibinfo {author} {\bibfnamefont
  {P.}~\bibnamefont {Addesso}}, \bibinfo {author} {\bibfnamefont {R.~X.}\
  \bibnamefont {Adhikari}}, \bibinfo {author} {\bibfnamefont {V.~B.}\
  \bibnamefont {Adya}}, \bibinfo {author} {\bibfnamefont {C.}~\bibnamefont
  {Affeldt}}, \bibinfo {author} {\bibfnamefont {M.}~\bibnamefont {Agathos}},
  \bibinfo {author} {\bibfnamefont {K.}~\bibnamefont {Agatsuma}}, \bibinfo
  {author} {\bibfnamefont {N.}~\bibnamefont {Aggarwal}}, \bibinfo {author}
  {\bibfnamefont {O.~D.}\ \bibnamefont {Aguiar}}, \ and\ \bibinfo {author}
  {\bibfnamefont {L.}~\bibnamefont {Aiello}}} (\bibinfo {year}
  {2017}{\natexlab{a}}),\ \bibfield  {title} {\enquote {\bibinfo {title} {First
  search for gravitational waves from known pulsars with advanced ligo},}\
  }\href {http://stacks.iop.org/0004-637X/839/i=1/a=12} {\bibfield  {journal}
  {\bibinfo  {journal} {The Astrophysical Journal}\ }\textbf {\bibinfo {volume}
  {839}}~(\bibinfo {number} {1}),\ \bibinfo {pages} {12}}\BibitemShut {NoStop}%
\bibitem [{\citenamefont {Abbott}\ \emph
  {et~al.}(2017{\natexlab{b}})\citenamefont {Abbott} \emph
  {et~al.}}]{Abbott:2017mwl}%
  \BibitemOpen
  \bibfield  {author} {\bibinfo {author} {\bibnamefont {Abbott}, \bibfnamefont
  {B~P}},  \emph {et~al.} (\bibinfo {collaboration} {Virgo, LIGO Scientific})}
  (\bibinfo {year} {2017}{\natexlab{b}}),\ \bibfield  {title} {\enquote
  {\bibinfo {title} {{Upper Limits on Gravitational Waves from Scorpius X-1
  from a Model-Based Cross-Correlation Search in Advanced LIGO Data}},}\
  }\href@noop {} {\ }\Eprint {http://arxiv.org/abs/1706.03119}
  {arXiv:1706.03119 [astro-ph.HE]} \BibitemShut {NoStop}%
\bibitem [{\citenamefont {Abbott}(2016)}]{PhysRevLett.116.061102}%
  \BibitemOpen
  \bibfield  {author} {\bibinfo {author} {\bibnamefont {Abbott}, \bibfnamefont
  {B~P et~al}} (\bibinfo {collaboration} {LIGO Scientific Collaboration and
  Virgo Collaboration})} (\bibinfo {year} {2016}),\ \bibfield  {title}
  {\enquote {\bibinfo {title} {Observation of gravitational waves from a binary
  black hole merger},}\ }\href {\doibase 10.1103/PhysRevLett.116.061102}
  {\bibfield  {journal} {\bibinfo  {journal} {Phys. Rev. Lett.}\ }\textbf
  {\bibinfo {volume} {116}},\ \bibinfo {pages} {061102}}\BibitemShut {NoStop}%
\bibitem [{\citenamefont {Adams}(1914)}]{Adams1914}%
  \BibitemOpen
  \bibfield  {author} {\bibinfo {author} {\bibnamefont {Adams}, \bibfnamefont
  {Walter~S}}} (\bibinfo {year} {1914}),\ \href@noop {} {\bibfield  {journal}
  {\bibinfo  {journal} {Pubs. of the Astron. Society of the Pacific}\ }\textbf
  {\bibinfo {volume} {26}},\ \bibinfo {pages} {198}}\BibitemShut {NoStop}%
\bibitem [{\citenamefont {Adams}(1915)}]{Adams1915}%
  \BibitemOpen
  \bibfield  {author} {\bibinfo {author} {\bibnamefont {Adams}, \bibfnamefont
  {Walter~S}}} (\bibinfo {year} {1915}),\ \href@noop {} {\bibfield  {journal}
  {\bibinfo  {journal} {Pubs. of the Astron. Society of the Pacific}\ }\textbf
  {\bibinfo {volume} {27}},\ \bibinfo {pages} {236}}\BibitemShut {NoStop}%
\bibitem [{\citenamefont {Aichelin}\ and\ \citenamefont
  {Stöcker}(1986)}]{AICHELIN198614}%
  \BibitemOpen
  \bibfield  {author} {\bibinfo {author} {\bibnamefont {Aichelin},
  \bibfnamefont {J}}, \ and\ \bibinfo {author} {\bibfnamefont {H.}~\bibnamefont
  {Stöcker}}} (\bibinfo {year} {1986}),\ \bibfield  {title} {\enquote
  {\bibinfo {title} {Quantum molecular dynamics ---€" a novel approach to
  n-body correlations in heavy ion collisions},}\ }\href {\doibase
  http://dx.doi.org/10.1016/0370-2693(86)90916-0} {\bibfield  {journal}
  {\bibinfo  {journal} {Physics Letters B}\ }\textbf {\bibinfo {volume}
  {176}}~(\bibinfo {number} {1}),\ \bibinfo {pages} {14 -- 19}}\BibitemShut
  {NoStop}%
\bibitem [{\citenamefont {Aasi~et al.}(2016)}]{arXiv:1605.03233}%
  \BibitemOpen
  \bibfield  {author} {\bibinfo {author} {\bibnamefont {Aasi~et al.},
  \bibfnamefont {J}}} (\bibinfo {year} {2016}),\ \bibfield  {title} {\enquote
  {\bibinfo {title} {Comprehensive all-sky search for periodic gravitational
  waves in the sixth science run ligo data},}\ }\href@noop {} {\bibinfo
  {journal} {arXiv:1605.03233}\ }\BibitemShut {NoStop}%
\bibitem [{\citenamefont {Alcain}\ \emph {et~al.}(2014)\citenamefont {Alcain},
  \citenamefont {Gim\'enez~Molinelli},\ and\ \citenamefont
  {Dorso}}]{PhysRevC.90.065803}%
  \BibitemOpen
\bibfield  {journal} {  }\bibfield  {author} {\bibinfo {author} {\bibnamefont
  {Alcain}, \bibfnamefont {P~N}}, \bibinfo {author} {\bibfnamefont {P.~A.}\
  \bibnamefont {Gim\'enez~Molinelli}}, \ and\ \bibinfo {author} {\bibfnamefont
  {C.~O.}\ \bibnamefont {Dorso}}} (\bibinfo {year} {2014}),\ \bibfield  {title}
  {\enquote {\bibinfo {title} {Beyond nuclear ``pasta'' : Phase transitions and
  neutrino opacity of new ``pasta'' phases},}\ }\href {\doibase
  10.1103/PhysRevC.90.065803} {\bibfield  {journal} {\bibinfo  {journal} {Phys.
  Rev. C}\ }\textbf {\bibinfo {volume} {90}},\ \bibinfo {pages}
  {065803}}\BibitemShut {NoStop}%
\bibitem [{\citenamefont {Andersson}\ \emph {et~al.}(1999)\citenamefont
  {Andersson}, \citenamefont {Kokkotas},\ and\ \citenamefont
  {Schutz}}]{0004-637X-510-2-846}%
  \BibitemOpen
  \bibfield  {author} {\bibinfo {author} {\bibnamefont {Andersson},
  \bibfnamefont {Nils}}, \bibinfo {author} {\bibfnamefont {Kostas}\
  \bibnamefont {Kokkotas}}, \ and\ \bibinfo {author} {\bibfnamefont
  {Bernard~F.}\ \bibnamefont {Schutz}}} (\bibinfo {year} {1999}),\ \bibfield
  {title} {\enquote {\bibinfo {title} {Gravitational radiation limit on the
  spin of young neutron stars},}\ }\href
  {http://stacks.iop.org/0004-637X/510/i=2/a=846} {\bibfield  {journal}
  {\bibinfo  {journal} {The Astrophysical Journal}\ }\textbf {\bibinfo {volume}
  {510}}~(\bibinfo {number} {2}),\ \bibinfo {pages} {846}}\BibitemShut
  {NoStop}%
\bibitem [{\citenamefont {Arzoumanian}\ \emph {et~al.}(2014)\citenamefont
  {Arzoumanian}, \citenamefont {Gendreau}, \citenamefont {Baker}, \citenamefont
  {Cazeau}, \citenamefont {Hestnes}, \citenamefont {Kellogg}, \citenamefont
  {Kenyon}, \citenamefont {Kozon}, \citenamefont {Liu}, \citenamefont
  {Manthripragada}, \citenamefont {Markwardt}, \citenamefont {Mitchell},
  \citenamefont {Mitchell}, \citenamefont {Monroe}, \citenamefont {Okajima},
  \citenamefont {Pollard}, \citenamefont {Powers}, \citenamefont {Savadkin},
  \citenamefont {Winternitz}, \citenamefont {Chen}, \citenamefont {Wright},
  \citenamefont {Foster}, \citenamefont {Prigozhin}, \citenamefont
  {Remillard},\ and\ \citenamefont {Doty}}]{NICER}%
  \BibitemOpen
  \bibfield  {author} {\bibinfo {author} {\bibnamefont {Arzoumanian},
  \bibfnamefont {Z}}, \bibinfo {author} {\bibfnamefont {K.~C.}\ \bibnamefont
  {Gendreau}}, \bibinfo {author} {\bibfnamefont {C.~L.}\ \bibnamefont {Baker}},
  \bibinfo {author} {\bibfnamefont {T.}~\bibnamefont {Cazeau}}, \bibinfo
  {author} {\bibfnamefont {P.}~\bibnamefont {Hestnes}}, \bibinfo {author}
  {\bibfnamefont {J.~W.}\ \bibnamefont {Kellogg}}, \bibinfo {author}
  {\bibfnamefont {S.~J.}\ \bibnamefont {Kenyon}}, \bibinfo {author}
  {\bibfnamefont {R.~P.}\ \bibnamefont {Kozon}}, \bibinfo {author}
  {\bibfnamefont {K.-C.}\ \bibnamefont {Liu}}, \bibinfo {author} {\bibfnamefont
  {S.~S.}\ \bibnamefont {Manthripragada}}, \bibinfo {author} {\bibfnamefont
  {C.~B.}\ \bibnamefont {Markwardt}}, \bibinfo {author} {\bibfnamefont {A.~L.}\
  \bibnamefont {Mitchell}}, \bibinfo {author} {\bibfnamefont {J.~W.}\
  \bibnamefont {Mitchell}}, \bibinfo {author} {\bibfnamefont {C.~A.}\
  \bibnamefont {Monroe}}, \bibinfo {author} {\bibfnamefont {T.}~\bibnamefont
  {Okajima}}, \bibinfo {author} {\bibfnamefont {S.~E.}\ \bibnamefont
  {Pollard}}, \bibinfo {author} {\bibfnamefont {D.~F.}\ \bibnamefont {Powers}},
  \bibinfo {author} {\bibfnamefont {B.~J.}\ \bibnamefont {Savadkin}}, \bibinfo
  {author} {\bibfnamefont {L.~B.}\ \bibnamefont {Winternitz}}, \bibinfo
  {author} {\bibfnamefont {P.~T.}\ \bibnamefont {Chen}}, \bibinfo {author}
  {\bibfnamefont {M.~R.}\ \bibnamefont {Wright}}, \bibinfo {author}
  {\bibfnamefont {R.}~\bibnamefont {Foster}}, \bibinfo {author} {\bibfnamefont
  {G.}~\bibnamefont {Prigozhin}}, \bibinfo {author} {\bibfnamefont
  {R.}~\bibnamefont {Remillard}}, \ and\ \bibinfo {author} {\bibfnamefont
  {J.}~\bibnamefont {Doty}}} (\bibinfo {year} {2014}),\ \bibfield  {title}
  {\enquote {\bibinfo {title} {The neutron star interior composition explorer
  (nicer): mission definition},}\ }\href {\doibase 10.1117/12.2056811}
  {\bibfield  {journal} {\bibinfo  {journal} {Proc. SPIE}\ }\textbf {\bibinfo
  {volume} {9144}},\ \bibinfo {pages} {914420--914420--9}}\BibitemShut
  {NoStop}%
\bibitem [{\citenamefont {Ball}(2014)}]{Ball2014}%
  \BibitemOpen
  \bibfield  {author} {\bibinfo {author} {\bibnamefont {Ball}, \bibfnamefont
  {Philip}}} (\bibinfo {year} {2014}),\ \href@noop {} {\bibfield  {journal}
  {\bibinfo  {journal} {Nature Materials}\ }\textbf {\bibinfo {volume} {13}},\
  \bibinfo {pages} {431}}\BibitemShut {NoStop}%
\bibitem [{\citenamefont {Baym}\ \emph {et~al.}(1971)\citenamefont {Baym},
  \citenamefont {Bethe},\ and\ \citenamefont {Pethick}}]{BAYM1971225}%
  \BibitemOpen
  \bibfield  {author} {\bibinfo {author} {\bibnamefont {Baym}, \bibfnamefont
  {Gordon}}, \bibinfo {author} {\bibfnamefont {Hans~A.}\ \bibnamefont {Bethe}},
  \ and\ \bibinfo {author} {\bibfnamefont {Christopher~J}\ \bibnamefont
  {Pethick}}} (\bibinfo {year} {1971}),\ \bibfield  {title} {\enquote {\bibinfo
  {title} {Neutron star matter},}\ }\href {\doibase
  http://dx.doi.org/10.1016/0375-9474(71)90281-8} {\bibfield  {journal}
  {\bibinfo  {journal} {Nuclear Physics A}\ }\textbf {\bibinfo {volume}
  {175}}~(\bibinfo {number} {2}),\ \bibinfo {pages} {225 -- 271}}\BibitemShut
  {NoStop}%
\bibitem [{\citenamefont {Berry}\ \emph {et~al.}(2016)\citenamefont {Berry},
  \citenamefont {Caplan}, \citenamefont {Horowitz}, \citenamefont {Huber},\
  and\ \citenamefont {Schneider}}]{Horowitz:2015gda}%
  \BibitemOpen
  \bibfield  {author} {\bibinfo {author} {\bibnamefont {Berry}, \bibfnamefont
  {D~K}}, \bibinfo {author} {\bibfnamefont {M.~E.}\ \bibnamefont {Caplan}},
  \bibinfo {author} {\bibfnamefont {C.~J.}\ \bibnamefont {Horowitz}}, \bibinfo
  {author} {\bibfnamefont {Greg}\ \bibnamefont {Huber}}, \ and\ \bibinfo
  {author} {\bibfnamefont {A.~S.}\ \bibnamefont {Schneider}}} (\bibinfo {year}
  {2016}),\ \bibfield  {title} {\enquote {\bibinfo {title} {{Parking-garage
  structures in nuclear astrophysics and cellular biophysics}},}\ }\href@noop
  {} {\bibfield  {journal} {\bibinfo  {journal} {Phys. Rev. C}\ }\textbf
  {\bibinfo {volume} {94}},\ \bibinfo {pages} {055801}},\ \Eprint
  {http://arxiv.org/abs/1509.00410} {arXiv:1509.00410 [nucl-th]} \BibitemShut
  {NoStop}%
\bibitem [{\citenamefont {Bildsten}(1998)}]{1538-4357-501-1-L89}%
  \BibitemOpen
  \bibfield  {author} {\bibinfo {author} {\bibnamefont {Bildsten},
  \bibfnamefont {Lars}}} (\bibinfo {year} {1998}),\ \bibfield  {title}
  {\enquote {\bibinfo {title} {Gravitational radiation and rotation of
  accreting neutron stars},}\ }\href
  {http://stacks.iop.org/1538-4357/501/i=1/a=L89} {\bibfield  {journal}
  {\bibinfo  {journal} {The Astrophysical Journal Letters}\ }\textbf {\bibinfo
  {volume} {501}}~(\bibinfo {number} {1}),\ \bibinfo {pages} {L89}}\BibitemShut
  {NoStop}%
\bibitem [{\citenamefont {Brown}\ and\ \citenamefont
  {Cumming}(2009)}]{crustcooling4}%
  \BibitemOpen
  \bibfield  {author} {\bibinfo {author} {\bibnamefont {Brown}, \bibfnamefont
  {Edward~F}}, \ and\ \bibinfo {author} {\bibfnamefont {Andrew}\ \bibnamefont
  {Cumming}}} (\bibinfo {year} {2009}),\ \href@noop {} {\bibinfo  {journal}
  {arXiv:0901.3115}\ }\BibitemShut {NoStop}%
\bibitem [{\citenamefont {Brush}\ \emph {et~al.}(1966)\citenamefont {Brush},
  \citenamefont {Sahlin},\ and\ \citenamefont {Teller}}]{jchemphys1966}%
  \BibitemOpen
\bibfield  {journal} {  }\bibfield  {author} {\bibinfo {author} {\bibnamefont
  {Brush}, \bibfnamefont {S~G}}, \bibinfo {author} {\bibfnamefont {H.~L.}\
  \bibnamefont {Sahlin}}, \ and\ \bibinfo {author} {\bibfnamefont
  {E.}~\bibnamefont {Teller}}} (\bibinfo {year} {1966}),\ \bibfield  {title}
  {\enquote {\bibinfo {title} {Monte carlo study of a one‐component plasma.
  i},}\ }\href {\doibase 10.1063/1.1727895} {\bibfield  {journal} {\bibinfo
  {journal} {The Journal of Chemical Physics}\ }\textbf {\bibinfo {volume}
  {45}}~(\bibinfo {number} {6}),\ \bibinfo {pages} {2102--2118}},\ \Eprint
  {http://arxiv.org/abs/http://aip.scitation.org/doi/pdf/10.1063/1.1727895}
  {http://aip.scitation.org/doi/pdf/10.1063/1.1727895} \BibitemShut {NoStop}%
\bibitem [{\citenamefont {Buehler}(2015)}]{buehler2015cell}%
  \BibitemOpen
  \bibfield  {author} {\bibinfo {author} {\bibnamefont {Buehler}, \bibfnamefont
  {L}}} (\bibinfo {year} {2015}),\ \href
  {https://books.google.com/books?id=j\_0pCgAAQBAJ} {\emph {\bibinfo {title}
  {Cell Membranes:}}}\ (\bibinfo  {publisher} {Taylor \& Francis
  Group})\BibitemShut {NoStop}%
\bibitem [{\citenamefont {Cackett}\ and\ \citenamefont
  {et~al.}(2006)}]{crustcooling1}%
  \BibitemOpen
  \bibfield  {author} {\bibinfo {author} {\bibnamefont {Cackett}, \bibfnamefont
  {E~M}}, \ and\ \bibinfo {author} {\bibnamefont {et~al.}}} (\bibinfo {year}
  {2006}),\ \href@noop {} {\bibfield  {journal} {\bibinfo  {journal} {Mon. Not.
  Roy. Astron. Soc.}\ }\textbf {\bibinfo {volume} {372}},\ \bibinfo {pages}
  {479}}\BibitemShut {NoStop}%
\bibitem [{\citenamefont {Camisassa}\ \emph {et~al.}(2017)\citenamefont
  {Camisassa}, \citenamefont {Althaus}, \citenamefont {Rohrmann}, \citenamefont
  {Garci­a-Berro}, \citenamefont {Torres}, \citenamefont {Corsico},\ and\
  \citenamefont {Wachlin}}]{0004-637X-839-1-11}%
  \BibitemOpen
  \bibfield  {author} {\bibinfo {author} {\bibnamefont {Camisassa},
  \bibfnamefont {Mari­a~E}}, \bibinfo {author} {\bibfnamefont {Leandro~G.}\
  \bibnamefont {Althaus}}, \bibinfo {author} {\bibfnamefont {Rene~D.}\
  \bibnamefont {Rohrmann}}, \bibinfo {author} {\bibfnamefont {Enrique}\
  \bibnamefont {Garci­a-Berro}}, \bibinfo {author} {\bibfnamefont {Santiago}\
  \bibnamefont {Torres}}, \bibinfo {author} {\bibfnamefont {Alejandro~H.}\
  \bibnamefont {Corsico}}, \ and\ \bibinfo {author} {\bibfnamefont {Felipe~C.}\
  \bibnamefont {Wachlin}}} (\bibinfo {year} {2017}),\ \bibfield  {title}
  {\enquote {\bibinfo {title} {Updated evolutionary sequences for
  hydrogen-deficient white dwarfs},}\ }\href
  {http://stacks.iop.org/0004-637X/839/i=1/a=11} {\bibfield  {journal}
  {\bibinfo  {journal} {The Astrophysical Journal}\ }\textbf {\bibinfo {volume}
  {839}}~(\bibinfo {number} {1}),\ \bibinfo {pages} {11}}\BibitemShut {NoStop}%
\bibitem [{\citenamefont {Caplan}\ \emph {et~al.}(2015)\citenamefont {Caplan},
  \citenamefont {Schneider}, \citenamefont {Horowitz},\ and\ \citenamefont
  {Berry}}]{PhysRevC.91.065802}%
  \BibitemOpen
  \bibfield  {author} {\bibinfo {author} {\bibnamefont {Caplan}, \bibfnamefont
  {M~E}}, \bibinfo {author} {\bibfnamefont {A.~S.}\ \bibnamefont {Schneider}},
  \bibinfo {author} {\bibfnamefont {C.~J.}\ \bibnamefont {Horowitz}}, \ and\
  \bibinfo {author} {\bibfnamefont {D.~K.}\ \bibnamefont {Berry}}} (\bibinfo
  {year} {2015}),\ \bibfield  {title} {\enquote {\bibinfo {title} {Pasta
  nucleosynthesis: Molecular dynamics simulations of nuclear statistical
  equilibrium},}\ }\href {\doibase 10.1103/PhysRevC.91.065802} {\bibfield
  {journal} {\bibinfo  {journal} {Phys. Rev. C}\ }\textbf {\bibinfo {volume}
  {91}},\ \bibinfo {pages} {065802}}\BibitemShut {NoStop}%
\bibitem [{\citenamefont {Chamel}(2005)}]{CHAMEL2005109}%
  \BibitemOpen
  \bibfield  {author} {\bibinfo {author} {\bibnamefont {Chamel}, \bibfnamefont
  {Nicolas}}} (\bibinfo {year} {2005}),\ \bibfield  {title} {\enquote {\bibinfo
  {title} {Band structure effects for dripped neutrons in neutron star
  crust},}\ }\href {\doibase http://dx.doi.org/10.1016/j.nuclphysa.2004.09.011}
  {\bibfield  {journal} {\bibinfo  {journal} {Nuclear Physics A}\ }\textbf
  {\bibinfo {volume} {747}}~(\bibinfo {number} {1}),\ \bibinfo {pages} {109 --
  128}}\BibitemShut {NoStop}%
\bibitem [{\citenamefont {{Chugunov}}\ and\ \citenamefont
  {{Horowitz}}(2010)}]{chugunov}%
  \BibitemOpen
  \bibfield  {author} {\bibinfo {author} {\bibnamefont {{Chugunov}},
  \bibfnamefont {A~I}}, \ and\ \bibinfo {author} {\bibfnamefont {C.~J.}\
  \bibnamefont {{Horowitz}}}} (\bibinfo {year} {2010}),\ \bibfield  {title}
  {\enquote {\bibinfo {title} {{Breaking stress of neutron star crust}},}\
  }\href {\doibase 10.1111/j.1745-3933.2010.00903.x} {\bibfield  {journal}
  {\bibinfo  {journal} {Monthly Notices Royal Astro. Soc.}\ }\textbf {\bibinfo
  {volume} {407}},\ \bibinfo {pages} {L54--L58}}\BibitemShut {NoStop}%
\bibitem [{\citenamefont {Costantini}\ \emph {et~al.}(2004)\citenamefont
  {Costantini}, \citenamefont {Ianni},\ and\ \citenamefont
  {Vissani}}]{PhysRevD.70.043006}%
  \BibitemOpen
  \bibfield  {author} {\bibinfo {author} {\bibnamefont {Costantini},
  \bibfnamefont {Maria~Laura}}, \bibinfo {author} {\bibfnamefont {Aldo}\
  \bibnamefont {Ianni}}, \ and\ \bibinfo {author} {\bibfnamefont {Francesco}\
  \bibnamefont {Vissani}}} (\bibinfo {year} {2004}),\ \bibfield  {title}
  {\enquote {\bibinfo {title} {Sn1987a and the properties of the neutrino
  burst},}\ }\href@noop {} {\bibfield  {journal} {\bibinfo  {journal} {Phys.
  Rev. D}\ }\textbf {\bibinfo {volume} {70}},\ \bibinfo {pages}
  {043006}}\BibitemShut {NoStop}%
\bibitem [{\citenamefont {Dorso}\ \emph {et~al.}(1987)\citenamefont {Dorso},
  \citenamefont {Duarte},\ and\ \citenamefont {Randrup}}]{DORSO1987287}%
  \BibitemOpen
  \bibfield  {author} {\bibinfo {author} {\bibnamefont {Dorso}, \bibfnamefont
  {Claudio}}, \bibinfo {author} {\bibfnamefont {Sergio}\ \bibnamefont
  {Duarte}}, \ and\ \bibinfo {author} {\bibfnamefont {Jørgen}\ \bibnamefont
  {Randrup}}} (\bibinfo {year} {1987}),\ \bibfield  {title} {\enquote {\bibinfo
  {title} {Classical simulation of the fermi gas},}\ }\href {\doibase
  http://dx.doi.org/10.1016/0370-2693(87)91382-7} {\bibfield  {journal}
  {\bibinfo  {journal} {Physics Letters B}\ }\textbf {\bibinfo {volume}
  {188}}~(\bibinfo {number} {3}),\ \bibinfo {pages} {287 -- 294}}\BibitemShut
  {NoStop}%
\bibitem [{\citenamefont {Engstrom}\ \emph {et~al.}(2016)\citenamefont
  {Engstrom}, \citenamefont {Yoder},\ and\ \citenamefont {Crespi}}]{Engstrom}%
  \BibitemOpen
  \bibfield  {author} {\bibinfo {author} {\bibnamefont {Engstrom},
  \bibfnamefont {T~A}}, \bibinfo {author} {\bibfnamefont {N.~C.}\ \bibnamefont
  {Yoder}}, \ and\ \bibinfo {author} {\bibfnamefont {V.~H.}\ \bibnamefont
  {Crespi}}} (\bibinfo {year} {2016}),\ \bibfield  {title} {\enquote {\bibinfo
  {title} {Crystal chemistry of three-component white dwarfs and neutron star
  crusts: Phase stability, phase stratification, and physical properties},}\
  }\href {http://stacks.iop.org/0004-637X/818/i=2/a=183} {\bibfield  {journal}
  {\bibinfo  {journal} {The Astrophysical Journal}\ }\textbf {\bibinfo {volume}
  {818}}~(\bibinfo {number} {2}),\ \bibinfo {pages} {183}}\BibitemShut
  {NoStop}%
\bibitem [{\citenamefont {Fontaine}\ \emph {et~al.}(2001)\citenamefont
  {Fontaine}, \citenamefont {Brassard},\ and\ \citenamefont
  {Bergeron}}]{cosmochron}%
  \BibitemOpen
  \bibfield  {author} {\bibinfo {author} {\bibnamefont {Fontaine},
  \bibfnamefont {G}}, \bibinfo {author} {\bibfnamefont {P.}~\bibnamefont
  {Brassard}}, \ and\ \bibinfo {author} {\bibfnamefont {P.}~\bibnamefont
  {Bergeron}}} (\bibinfo {year} {2001}),\ \href@noop {} {\bibfield  {journal}
  {\bibinfo  {journal} {PASP}\ }\textbf {\bibinfo {volume} {113}},\ \bibinfo
  {pages} {409}}\BibitemShut {NoStop}%
\bibitem [{\citenamefont {Garcia-Berro}\ \emph {et~al.}(2011)\citenamefont
  {Garcia-Berro}, \citenamefont {Torres}, \citenamefont {Renedo}, \citenamefont
  {Camacho}, \citenamefont {Althaus}, \citenamefont {Corsico}, \citenamefont
  {Salaris},\ and\ \citenamefont {Isern}}]{garcia-berro}%
  \BibitemOpen
  \bibfield  {author} {\bibinfo {author} {\bibnamefont {Garcia-Berro},
  \bibfnamefont {E}}, \bibinfo {author} {\bibfnamefont {S.}~\bibnamefont
  {Torres}}, \bibinfo {author} {\bibfnamefont {I.}~\bibnamefont {Renedo}},
  \bibinfo {author} {\bibfnamefont {J.}~\bibnamefont {Camacho}}, \bibinfo
  {author} {\bibfnamefont {L.~G.}\ \bibnamefont {Althaus}}, \bibinfo {author}
  {\bibfnamefont {A.~H.}\ \bibnamefont {Corsico}}, \bibinfo {author}
  {\bibfnamefont {M.}~\bibnamefont {Salaris}}, \ and\ \bibinfo {author}
  {\bibfnamefont {J.}~\bibnamefont {Isern}}} (\bibinfo {year} {2011}),\
  \href@noop {} {\bibinfo  {journal} {arXiv:1107.3016}\ }\BibitemShut {NoStop}%
\bibitem [{\citenamefont {G\"ogelein}\ and\ \citenamefont
  {M\"uther}(2007)}]{PhysRevC.76.024312}%
  \BibitemOpen
\bibfield  {journal} {  }\bibfield  {author} {\bibinfo {author} {\bibnamefont
  {G\"ogelein}, \bibfnamefont {P}}, \ and\ \bibinfo {author} {\bibfnamefont
  {H.}~\bibnamefont {M\"uther}}} (\bibinfo {year} {2007}),\ \bibfield  {title}
  {\enquote {\bibinfo {title} {Nuclear matter in the crust of neutron stars},}\
  }\href {\doibase 10.1103/PhysRevC.76.024312} {\bibfield  {journal} {\bibinfo
  {journal} {Phys. Rev. C}\ }\textbf {\bibinfo {volume} {76}},\ \bibinfo
  {pages} {024312}}\BibitemShut {NoStop}%
\bibitem [{\citenamefont {Gupta}\ \emph {et~al.}(2007)\citenamefont {Gupta},
  \citenamefont {Brown}, \citenamefont {Schatz}, \citenamefont {Moller},\ and\
  \citenamefont {Kratz}}]{rpash3}%
  \BibitemOpen
  \bibfield  {author} {\bibinfo {author} {\bibnamefont {Gupta}, \bibfnamefont
  {S}}, \bibinfo {author} {\bibfnamefont {E.~F.}\ \bibnamefont {Brown}},
  \bibinfo {author} {\bibfnamefont {H.}~\bibnamefont {Schatz}}, \bibinfo
  {author} {\bibfnamefont {P.}~\bibnamefont {Moller}}, \ and\ \bibinfo {author}
  {\bibfnamefont {K-L.}\ \bibnamefont {Kratz}}} (\bibinfo {year} {2007}),\
  \href@noop {} {\bibfield  {journal} {\bibinfo  {journal} {Astrophys. J.}\
  }\textbf {\bibinfo {volume} {662}},\ \bibinfo {pages} {1188}}\BibitemShut
  {NoStop}%
\bibitem [{\citenamefont {{Hashimoto}}\ \emph {et~al.}(1984)\citenamefont
  {{Hashimoto}}, \citenamefont {{Seki}},\ and\ \citenamefont
  {{Yamada}}}]{1984PThPh..71..320H}%
  \BibitemOpen
  \bibfield  {author} {\bibinfo {author} {\bibnamefont {{Hashimoto}},
  \bibfnamefont {M}}, \bibinfo {author} {\bibfnamefont {H.}~\bibnamefont
  {{Seki}}}, \ and\ \bibinfo {author} {\bibfnamefont {M.}~\bibnamefont
  {{Yamada}}}} (\bibinfo {year} {1984}),\ \bibfield  {title} {\enquote
  {\bibinfo {title} {{Shape of nuclei in the crust of a neutron star.}}}\
  }\href {\doibase 10.1143/PTP.71.320} {\bibfield  {journal} {\bibinfo
  {journal} {Progress of Theoretical Physics}\ }\textbf {\bibinfo {volume}
  {71}},\ \bibinfo {pages} {320--326}}\BibitemShut {NoStop}%
\bibitem [{\citenamefont {Horowitz}(2010)}]{lowmassNS}%
  \BibitemOpen
  \bibfield  {author} {\bibinfo {author} {\bibnamefont {Horowitz},
  \bibfnamefont {C~J}}} (\bibinfo {year} {2010}),\ \bibfield  {title} {\enquote
  {\bibinfo {title} {Gravitational waves from low mass neutron stars},}\ }\href
  {\doibase 10.1103/PhysRevD.81.103001} {\bibfield  {journal} {\bibinfo
  {journal} {Phys. Rev. D}\ }\textbf {\bibinfo {volume} {81}},\ \bibinfo
  {pages} {103001}}\BibitemShut {NoStop}%
\bibitem [{\citenamefont {Horowitz}\ and\ \citenamefont
  {Berry}(2008)}]{PhysRevC.78.035806}%
  \BibitemOpen
  \bibfield  {author} {\bibinfo {author} {\bibnamefont {Horowitz},
  \bibfnamefont {C~J}}, \ and\ \bibinfo {author} {\bibfnamefont {D.~K.}\
  \bibnamefont {Berry}}} (\bibinfo {year} {2008}),\ \bibfield  {title}
  {\enquote {\bibinfo {title} {Shear viscosity and thermal conductivity of
  nuclear ``pasta''},}\ }\href {\doibase 10.1103/PhysRevC.78.035806} {\bibfield
   {journal} {\bibinfo  {journal} {Phys. Rev. C}\ }\textbf {\bibinfo {volume}
  {78}},\ \bibinfo {pages} {035806}}\BibitemShut {NoStop}%
\bibitem [{\citenamefont {Horowitz}\ \emph {et~al.}(2015)\citenamefont
  {Horowitz}, \citenamefont {Berry}, \citenamefont {Briggs}, \citenamefont
  {Caplan}, \citenamefont {Cumming},\ and\ \citenamefont
  {Schneider}}]{PhysRevLett.114.031102}%
  \BibitemOpen
  \bibfield  {author} {\bibinfo {author} {\bibnamefont {Horowitz},
  \bibfnamefont {C~J}}, \bibinfo {author} {\bibfnamefont {D.~K.}\ \bibnamefont
  {Berry}}, \bibinfo {author} {\bibfnamefont {C.~M.}\ \bibnamefont {Briggs}},
  \bibinfo {author} {\bibfnamefont {M.~E.}\ \bibnamefont {Caplan}}, \bibinfo
  {author} {\bibfnamefont {A.}~\bibnamefont {Cumming}}, \ and\ \bibinfo
  {author} {\bibfnamefont {A.~S.}\ \bibnamefont {Schneider}}} (\bibinfo {year}
  {2015}),\ \bibfield  {title} {\enquote {\bibinfo {title} {Disordered nuclear
  pasta, magnetic field decay, and crust cooling in neutron stars},}\ }\href
  {\doibase 10.1103/PhysRevLett.114.031102} {\bibfield  {journal} {\bibinfo
  {journal} {Phys. Rev. Lett.}\ }\textbf {\bibinfo {volume} {114}},\ \bibinfo
  {pages} {031102}}\BibitemShut {NoStop}%
\bibitem [{\citenamefont {Horowitz}\ \emph {et~al.}(2007)\citenamefont
  {Horowitz}, \citenamefont {Berry},\ and\ \citenamefont {Brown}}]{phasesep}%
  \BibitemOpen
  \bibfield  {author} {\bibinfo {author} {\bibnamefont {Horowitz},
  \bibfnamefont {C~J}}, \bibinfo {author} {\bibfnamefont {D.~K.}\ \bibnamefont
  {Berry}}, \ and\ \bibinfo {author} {\bibfnamefont {E.~F.}\ \bibnamefont
  {Brown}}} (\bibinfo {year} {2007}),\ \href@noop {} {\bibfield  {journal}
  {\bibinfo  {journal} {Phys. Rev. E}\ }\textbf {\bibinfo {volume} {75}},\
  \bibinfo {pages} {066101}}\BibitemShut {NoStop}%
\bibitem [{\citenamefont {Horowitz}\ \emph {et~al.}(2016)\citenamefont
  {Horowitz}, \citenamefont {Berry}, \citenamefont {Caplan}, \citenamefont
  {Fischer}, \citenamefont {Lin}, \citenamefont {Newton}, \citenamefont
  {O'Connor},\ and\ \citenamefont {Roberts}}]{Horowitz:2016fpa}%
  \BibitemOpen
  \bibfield  {author} {\bibinfo {author} {\bibnamefont {Horowitz},
  \bibfnamefont {C~J}}, \bibinfo {author} {\bibfnamefont {D.~K.}\ \bibnamefont
  {Berry}}, \bibinfo {author} {\bibfnamefont {M.~E.}\ \bibnamefont {Caplan}},
  \bibinfo {author} {\bibfnamefont {T.}~\bibnamefont {Fischer}}, \bibinfo
  {author} {\bibfnamefont {Zidu}\ \bibnamefont {Lin}}, \bibinfo {author}
  {\bibfnamefont {W.~G.}\ \bibnamefont {Newton}}, \bibinfo {author}
  {\bibfnamefont {E.}~\bibnamefont {O'Connor}}, \ and\ \bibinfo {author}
  {\bibfnamefont {L.~F.}\ \bibnamefont {Roberts}}} (\bibinfo {year} {2016}),\
  \bibfield  {title} {\enquote {\bibinfo {title} {{Nuclear pasta and supernova
  neutrinos at late times}},}\ }\href@noop {} {\ }\Eprint
  {http://arxiv.org/abs/1611.10226} {arXiv:1611.10226 [astro-ph.HE]}
  \BibitemShut {NoStop}%
\bibitem [{\citenamefont {Horowitz}\ and\ \citenamefont
  {Kadau}(2009)}]{crustbreaking}%
  \BibitemOpen
  \bibfield  {author} {\bibinfo {author} {\bibnamefont {Horowitz},
  \bibfnamefont {C~J}}, \ and\ \bibinfo {author} {\bibfnamefont {Kai}\
  \bibnamefont {Kadau}}} (\bibinfo {year} {2009}),\ \bibfield  {title}
  {\enquote {\bibinfo {title} {Breaking strain of neutron star crust and
  gravitational waves},}\ }\href {\doibase 10.1103/PhysRevLett.102.191102}
  {\bibfield  {journal} {\bibinfo  {journal} {Phys. Rev. Lett.}\ }\textbf
  {\bibinfo {volume} {102}},\ \bibinfo {pages} {191102}}\BibitemShut {NoStop}%
\bibitem [{\citenamefont {Horowitz}\ \emph
  {et~al.}(2004{\natexlab{a}})\citenamefont {Horowitz}, \citenamefont
  {P\'erez-Garc\'{\i}a},\ and\ \citenamefont
  {Piekarewicz}}]{PhysRevC.69.045804}%
  \BibitemOpen
  \bibfield  {author} {\bibinfo {author} {\bibnamefont {Horowitz},
  \bibfnamefont {C~J}}, \bibinfo {author} {\bibfnamefont {M.~A.}\ \bibnamefont
  {P\'erez-Garc\'{\i}a}}, \ and\ \bibinfo {author} {\bibfnamefont
  {J.}~\bibnamefont {Piekarewicz}}} (\bibinfo {year} {2004}{\natexlab{a}}),\
  \bibfield  {title} {\enquote {\bibinfo {title} {Neutrino-``pasta''
  scattering: The opacity of nonuniform neutron-rich matter},}\ }\href
  {\doibase 10.1103/PhysRevC.69.045804} {\bibfield  {journal} {\bibinfo
  {journal} {Phys. Rev. C}\ }\textbf {\bibinfo {volume} {69}},\ \bibinfo
  {pages} {045804}}\BibitemShut {NoStop}%
\bibitem [{\citenamefont {Horowitz}\ \emph {et~al.}(2010)\citenamefont
  {Horowitz}, \citenamefont {Schneider},\ and\ \citenamefont {Berry}}]{WD_PRL}%
  \BibitemOpen
  \bibfield  {author} {\bibinfo {author} {\bibnamefont {Horowitz},
  \bibfnamefont {C~J}}, \bibinfo {author} {\bibfnamefont {A.~S.}\ \bibnamefont
  {Schneider}}, \ and\ \bibinfo {author} {\bibfnamefont {D.~K.}\ \bibnamefont
  {Berry}}} (\bibinfo {year} {2010}),\ \href@noop {} {\bibfield  {journal}
  {\bibinfo  {journal} {Phys. Rev. Lett.}\ }\textbf {\bibinfo {volume} {104}},\
  \bibinfo {pages} {231101}}\BibitemShut {NoStop}%
\bibitem [{\citenamefont {Horowitz}\ \emph
  {et~al.}(2004{\natexlab{b}})\citenamefont {Horowitz}, \citenamefont
  {Perez-Garcia},\ and\ \citenamefont {Piekarewicz}}]{Horowitz:2004yf}%
  \BibitemOpen
  \bibfield  {author} {\bibinfo {author} {\bibnamefont {Horowitz},
  \bibfnamefont {Charles~J}}, \bibinfo {author} {\bibfnamefont {M.~A.}\
  \bibnamefont {Perez-Garcia}}, \ and\ \bibinfo {author} {\bibfnamefont
  {J.}~\bibnamefont {Piekarewicz}}} (\bibinfo {year} {2004}{\natexlab{b}}),\
  \bibfield  {title} {\enquote {\bibinfo {title} {{Neutrino-pasta scattering:
  The opacity of nonuniform neutron-rich matter}},}\ }\href {\doibase
  10.1103/PhysRevC.69.045804} {\bibfield  {journal} {\bibinfo  {journal} {Phys.
  Rev.}\ }\textbf {\bibinfo {volume} {C69}},\ \bibinfo {pages} {045804}},\
  \Eprint {http://arxiv.org/abs/astro-ph/0401079} {arXiv:astro-ph/0401079}
  \BibitemShut {NoStop}%
\bibitem [{\citenamefont {Hughto}\ \emph {et~al.}(2012)\citenamefont {Hughto},
  \citenamefont {Horowitz}, \citenamefont {Schneider}, \citenamefont {Medin},
  \citenamefont {Cumming},\ and\ \citenamefont {Berry}}]{PhysRevE.86.066413}%
  \BibitemOpen
  \bibfield  {author} {\bibinfo {author} {\bibnamefont {Hughto}, \bibfnamefont
  {J}}, \bibinfo {author} {\bibfnamefont {C.~J.}\ \bibnamefont {Horowitz}},
  \bibinfo {author} {\bibfnamefont {A.~S.}\ \bibnamefont {Schneider}}, \bibinfo
  {author} {\bibfnamefont {Zach}\ \bibnamefont {Medin}}, \bibinfo {author}
  {\bibfnamefont {Andrew}\ \bibnamefont {Cumming}}, \ and\ \bibinfo {author}
  {\bibfnamefont {D.~K.}\ \bibnamefont {Berry}}} (\bibinfo {year} {2012}),\
  \bibfield  {title} {\enquote {\bibinfo {title} {Direct molecular dynamics
  simulation of liquid-solid phase equilibria for a three-component plasma},}\
  }\href {\doibase 10.1103/PhysRevE.86.066413} {\bibfield  {journal} {\bibinfo
  {journal} {Phys. Rev. E}\ }\textbf {\bibinfo {volume} {86}},\ \bibinfo
  {pages} {066413}}\BibitemShut {NoStop}%
\bibitem [{\citenamefont {Hughto}\ \emph {et~al.}(2011)\citenamefont {Hughto},
  \citenamefont {Schneider}, \citenamefont {Horowitz},\ and\ \citenamefont
  {Berry}}]{soliddiffusion}%
  \BibitemOpen
  \bibfield  {author} {\bibinfo {author} {\bibnamefont {Hughto}, \bibfnamefont
  {J}}, \bibinfo {author} {\bibfnamefont {A.~S.}\ \bibnamefont {Schneider}},
  \bibinfo {author} {\bibfnamefont {C.~J.}\ \bibnamefont {Horowitz}}, \ and\
  \bibinfo {author} {\bibfnamefont {D.~K.}\ \bibnamefont {Berry}}} (\bibinfo
  {year} {2011}),\ \href@noop {} {\bibfield  {journal} {\bibinfo  {journal}
  {Phys. Rev. E}\ }\textbf {\bibinfo {volume} {84}},\ \bibinfo {pages}
  {016401}}\BibitemShut {NoStop}%
\bibitem [{\citenamefont {Ichimaru}\ \emph {et~al.}(1988)\citenamefont
  {Ichimaru}, \citenamefont {Iyetomi},\ and\ \citenamefont
  {Ogata}}]{Ichimaru88}%
  \BibitemOpen
  \bibfield  {author} {\bibinfo {author} {\bibnamefont {Ichimaru},
  \bibfnamefont {S}}, \bibinfo {author} {\bibfnamefont {H.}~\bibnamefont
  {Iyetomi}}, \ and\ \bibinfo {author} {\bibfnamefont {S}~\bibnamefont
  {Ogata}}} (\bibinfo {year} {1988}),\ \href@noop {} {\bibfield  {journal}
  {\bibinfo  {journal} {Astrophys. J.}\ }\textbf {\bibinfo {volume} {334}},\
  \bibinfo {pages} {L17}}\BibitemShut {NoStop}%
\bibitem [{\citenamefont {Jones}\ and\ \citenamefont
  {Ceperley}(1996)}]{Jones96}%
  \BibitemOpen
  \bibfield  {author} {\bibinfo {author} {\bibnamefont {Jones}, \bibfnamefont
  {M~D}}, \ and\ \bibinfo {author} {\bibfnamefont {D.~M.}\ \bibnamefont
  {Ceperley}}} (\bibinfo {year} {1996}),\ \bibfield  {title} {\enquote
  {\bibinfo {title} {Crystallization of the one-component plasma at finite
  temperature},}\ }\href {\doibase 10.1103/PhysRevLett.76.4572} {\bibfield
  {journal} {\bibinfo  {journal} {Phys. Rev. Lett.}\ }\textbf {\bibinfo
  {volume} {76}},\ \bibinfo {pages} {4572--4575}}\BibitemShut {NoStop}%
\bibitem [{\citenamefont {Kob}\ and\ \citenamefont
  {Andersen}(1994)}]{PhysRevLett.73.1376}%
  \BibitemOpen
  \bibfield  {author} {\bibinfo {author} {\bibnamefont {Kob}, \bibfnamefont
  {Walter}}, \ and\ \bibinfo {author} {\bibfnamefont {Hans~C.}\ \bibnamefont
  {Andersen}}} (\bibinfo {year} {1994}),\ \bibfield  {title} {\enquote
  {\bibinfo {title} {Scaling behavior in the $\ensuremath{\beta}$-relaxation
  regime of a supercooled lennard-jones mixture},}\ }\href {\doibase
  10.1103/PhysRevLett.73.1376} {\bibfield  {journal} {\bibinfo  {journal}
  {Phys. Rev. Lett.}\ }\textbf {\bibinfo {volume} {73}},\ \bibinfo {pages}
  {1376--1379}}\BibitemShut {NoStop}%
\bibitem [{\citenamefont {Kobyakov}\ and\ \citenamefont
  {Pethick}(2014)}]{Kobyakov2014}%
  \BibitemOpen
  \bibfield  {author} {\bibinfo {author} {\bibnamefont {Kobyakov},
  \bibfnamefont {D}}, \ and\ \bibinfo {author} {\bibfnamefont {C.~J.}\
  \bibnamefont {Pethick}}} (\bibinfo {year} {2014}),\ \href@noop {} {\bibfield
  {journal} {\bibinfo  {journal} {Phys. Rev. Lett.}\ }\textbf {\bibinfo
  {volume} {112}},\ \bibinfo {pages} {112504}}\BibitemShut {NoStop}%
\bibitem [{\citenamefont {Kozhberov}\ and\ \citenamefont
  {Baiko}(2015)}]{doi:10.1063/1.4930215}%
  \BibitemOpen
  \bibfield  {author} {\bibinfo {author} {\bibnamefont {Kozhberov},
  \bibfnamefont {A~A}}, \ and\ \bibinfo {author} {\bibfnamefont {D.~A.}\
  \bibnamefont {Baiko}}} (\bibinfo {year} {2015}),\ \bibfield  {title}
  {\enquote {\bibinfo {title} {Coulomb crystal mixtures in white dwarf cores
  and neutron star crusts},}\ }\href {\doibase 10.1063/1.4930215} {\bibfield
  {journal} {\bibinfo  {journal} {Physics of Plasmas}\ }\textbf {\bibinfo
  {volume} {22}}~(\bibinfo {number} {9}),\ \bibinfo {pages} {092903}},\ \Eprint
  {http://arxiv.org/abs/http://dx.doi.org/10.1063/1.4930215}
  {http://dx.doi.org/10.1063/1.4930215} \BibitemShut {NoStop}%
\bibitem [{\citenamefont {Lamb}\ \emph {et~al.}(1978)\citenamefont {Lamb},
  \citenamefont {Lattimer}, \citenamefont {Pethick},\ and\ \citenamefont
  {Ravenhall}}]{PhysRevLett.41.1623}%
  \BibitemOpen
  \bibfield  {author} {\bibinfo {author} {\bibnamefont {Lamb}, \bibfnamefont
  {D~Q}}, \bibinfo {author} {\bibfnamefont {J.~M.}\ \bibnamefont {Lattimer}},
  \bibinfo {author} {\bibfnamefont {C.~J.}\ \bibnamefont {Pethick}}, \ and\
  \bibinfo {author} {\bibfnamefont {D.~G.}\ \bibnamefont {Ravenhall}}}
  (\bibinfo {year} {1978}),\ \bibfield  {title} {\enquote {\bibinfo {title}
  {Hot dense matter and stellar collapse},}\ }\href {\doibase
  10.1103/PhysRevLett.41.1623} {\bibfield  {journal} {\bibinfo  {journal}
  {Phys. Rev. Lett.}\ }\textbf {\bibinfo {volume} {41}},\ \bibinfo {pages}
  {1623--1626}}\BibitemShut {NoStop}%
\bibitem [{\citenamefont {Lattimer}\ \emph {et~al.}(1985)\citenamefont
  {Lattimer}, \citenamefont {Pethick}, \citenamefont {Ravenhall},\ and\
  \citenamefont {Lamb}}]{LATTIMER1985646}%
  \BibitemOpen
  \bibfield  {author} {\bibinfo {author} {\bibnamefont {Lattimer},
  \bibfnamefont {JM}}, \bibinfo {author} {\bibfnamefont {C.J.}\ \bibnamefont
  {Pethick}}, \bibinfo {author} {\bibfnamefont {D.G.}\ \bibnamefont
  {Ravenhall}}, \ and\ \bibinfo {author} {\bibfnamefont {D.Q.}\ \bibnamefont
  {Lamb}}} (\bibinfo {year} {1985}),\ \bibfield  {title} {\enquote {\bibinfo
  {title} {Physical properties of hot, dense matter: The general case},}\
  }\href {\doibase http://dx.doi.org/10.1016/0375-9474(85)90006-5} {\bibfield
  {journal} {\bibinfo  {journal} {Nuclear Physics A}\ }\textbf {\bibinfo
  {volume} {432}}~(\bibinfo {number} {3}),\ \bibinfo {pages} {646 --
  742}}\BibitemShut {NoStop}%
\bibitem [{\citenamefont {Lorenz}\ \emph {et~al.}(1993)\citenamefont {Lorenz},
  \citenamefont {Ravenhall},\ and\ \citenamefont
  {Pethick}}]{PhysRevLett.70.379}%
  \BibitemOpen
  \bibfield  {author} {\bibinfo {author} {\bibnamefont {Lorenz}, \bibfnamefont
  {C~P}}, \bibinfo {author} {\bibfnamefont {D.~G.}\ \bibnamefont {Ravenhall}},
  \ and\ \bibinfo {author} {\bibfnamefont {C.~J.}\ \bibnamefont {Pethick}}}
  (\bibinfo {year} {1993}),\ \bibfield  {title} {\enquote {\bibinfo {title}
  {Neutron star crusts},}\ }\href {\doibase 10.1103/PhysRevLett.70.379}
  {\bibfield  {journal} {\bibinfo  {journal} {Phys. Rev. Lett.}\ }\textbf
  {\bibinfo {volume} {70}},\ \bibinfo {pages} {379--382}}\BibitemShut {NoStop}%
\bibitem [{\citenamefont {Loumos}\ and\ \citenamefont
  {Hubbard}(1973)}]{Loumos73}%
  \BibitemOpen
  \bibfield  {author} {\bibinfo {author} {\bibnamefont {Loumos}, \bibfnamefont
  {Gregory~L}}, \ and\ \bibinfo {author} {\bibfnamefont {W.~B.}\ \bibnamefont
  {Hubbard}}} (\bibinfo {year} {1973}),\ \href@noop {} {\bibfield  {journal}
  {\bibinfo  {journal} {Astrophys. J.}\ }\textbf {\bibinfo {volume} {180}},\
  \bibinfo {pages} {199}}\BibitemShut {NoStop}%
\bibitem [{\citenamefont {Magierski}\ and\ \citenamefont
  {Heenen}(2002)}]{Magierski2002}%
  \BibitemOpen
  \bibfield  {author} {\bibinfo {author} {\bibnamefont {Magierski},
  \bibfnamefont {P}}, \ and\ \bibinfo {author} {\bibfnamefont {P.-H.}\
  \bibnamefont {Heenen}}} (\bibinfo {year} {2002}),\ \bibfield  {title}
  {\enquote {\bibinfo {title} {Structure of the inner crust of neutron stars:
  Crystal lattice or disordered phase?}}\ }\href {\doibase
  10.1103/PhysRevC.65.045804} {\bibfield  {journal} {\bibinfo  {journal} {Phys.
  Rev. C}\ }\textbf {\bibinfo {volume} {65}},\ \bibinfo {pages}
  {045804}}\BibitemShut {NoStop}%
\bibitem [{\citenamefont {Maruyama}\ \emph {et~al.}(1998)\citenamefont
  {Maruyama}, \citenamefont {Niita}, \citenamefont {Oyamatsu}, \citenamefont
  {Maruyama}, \citenamefont {Chiba},\ and\ \citenamefont
  {Iwamoto}}]{PhysRevC.57.655}%
  \BibitemOpen
  \bibfield  {author} {\bibinfo {author} {\bibnamefont {Maruyama},
  \bibfnamefont {Toshiki}}, \bibinfo {author} {\bibfnamefont {Koji}\
  \bibnamefont {Niita}}, \bibinfo {author} {\bibfnamefont {Kazuhiro}\
  \bibnamefont {Oyamatsu}}, \bibinfo {author} {\bibfnamefont {Tomoyuki}\
  \bibnamefont {Maruyama}}, \bibinfo {author} {\bibfnamefont {Satoshi}\
  \bibnamefont {Chiba}}, \ and\ \bibinfo {author} {\bibfnamefont {Akira}\
  \bibnamefont {Iwamoto}}} (\bibinfo {year} {1998}),\ \bibfield  {title}
  {\enquote {\bibinfo {title} {Quantum molecular dynamics approach to the
  nuclear matter below the saturation density},}\ }\href {\doibase
  10.1103/PhysRevC.57.655} {\bibfield  {journal} {\bibinfo  {journal} {Phys.
  Rev. C}\ }\textbf {\bibinfo {volume} {57}},\ \bibinfo {pages}
  {655--665}}\BibitemShut {NoStop}%
\bibitem [{\citenamefont {Mckinven}\ \emph {et~al.}(2016)\citenamefont
  {Mckinven}, \citenamefont {Cumming}, \citenamefont {Medin},\ and\
  \citenamefont {Schatz}}]{Mckinven:2016zkg}%
  \BibitemOpen
  \bibfield  {author} {\bibinfo {author} {\bibnamefont {Mckinven},
  \bibfnamefont {Ryan}}, \bibinfo {author} {\bibfnamefont {Andrew}\
  \bibnamefont {Cumming}}, \bibinfo {author} {\bibfnamefont {Zach}\
  \bibnamefont {Medin}}, \ and\ \bibinfo {author} {\bibfnamefont {Hendrik}\
  \bibnamefont {Schatz}}} (\bibinfo {year} {2016}),\ \bibfield  {title}
  {\enquote {\bibinfo {title} {A survey of chemical separation in accreting
  neutron stars},}\ }\href {http://stacks.iop.org/0004-637X/823/i=2/a=117}
  {\bibfield  {journal} {\bibinfo  {journal} {The Astrophysical Journal}\
  }\textbf {\bibinfo {volume} {823}}~(\bibinfo {number} {2}),\ \bibinfo {pages}
  {117}}\BibitemShut {NoStop}%
\bibitem [{\citenamefont {Meadors}\ \emph {et~al.}(2017)\citenamefont
  {Meadors}, \citenamefont {Goetz}, \citenamefont {Riles}, \citenamefont
  {Creighton},\ and\ \citenamefont {Robinet}}]{PhysRevD.95.042005}%
  \BibitemOpen
  \bibfield  {author} {\bibinfo {author} {\bibnamefont {Meadors}, \bibfnamefont
  {G~D}}, \bibinfo {author} {\bibfnamefont {E.}~\bibnamefont {Goetz}}, \bibinfo
  {author} {\bibfnamefont {K.}~\bibnamefont {Riles}}, \bibinfo {author}
  {\bibfnamefont {T.}~\bibnamefont {Creighton}}, \ and\ \bibinfo {author}
  {\bibfnamefont {F.}~\bibnamefont {Robinet}}} (\bibinfo {year} {2017}),\
  \bibfield  {title} {\enquote {\bibinfo {title} {Searches for continuous
  gravitational waves from scorpius x-1 and xte j1751-305 in ligo's sixth
  science run},}\ }\href {\doibase 10.1103/PhysRevD.95.042005} {\bibfield
  {journal} {\bibinfo  {journal} {Phys. Rev. D}\ }\textbf {\bibinfo {volume}
  {95}},\ \bibinfo {pages} {042005}}\BibitemShut {NoStop}%
\bibitem [{\citenamefont {Medin}\ and\ \citenamefont
  {Cumming}(2010)}]{Medin2010}%
  \BibitemOpen
  \bibfield  {author} {\bibinfo {author} {\bibnamefont {Medin}, \bibfnamefont
  {Zach}}, \ and\ \bibinfo {author} {\bibfnamefont {Andrew}\ \bibnamefont
  {Cumming}}} (\bibinfo {year} {2010}),\ \bibfield  {title} {\enquote {\bibinfo
  {title} {Crystallization of classical multicomponent plasmas},}\ }\href
  {\doibase 10.1103/PhysRevE.81.036107} {\bibfield  {journal} {\bibinfo
  {journal} {Phys. Rev. E}\ }\textbf {\bibinfo {volume} {81}},\ \bibinfo
  {pages} {036107}}\BibitemShut {NoStop}%
\bibitem [{\citenamefont {Medin}\ and\ \citenamefont
  {Cumming}(2014)}]{2041-8205-783-1-L3}%
  \BibitemOpen
  \bibfield  {author} {\bibinfo {author} {\bibnamefont {Medin}, \bibfnamefont
  {Zach}}, \ and\ \bibinfo {author} {\bibfnamefont {Andrew}\ \bibnamefont
  {Cumming}}} (\bibinfo {year} {2014}),\ \bibfield  {title} {\enquote {\bibinfo
  {title} {A signature of chemical separation in the cooling light curves of
  transiently accreting neutron stars},}\ }\href
  {http://stacks.iop.org/2041-8205/783/i=1/a=L3} {\bibfield  {journal}
  {\bibinfo  {journal} {The Astrophysical Journal Letters}\ }\textbf {\bibinfo
  {volume} {783}}~(\bibinfo {number} {1}),\ \bibinfo {pages} {L3}}\BibitemShut
  {NoStop}%
\bibitem [{\citenamefont {Medin}\ and\ \citenamefont
  {Cumming}(2015)}]{Medin:2014zfa}%
  \BibitemOpen
  \bibfield  {author} {\bibinfo {author} {\bibnamefont {Medin}, \bibfnamefont
  {Zach}}, \ and\ \bibinfo {author} {\bibfnamefont {Andrew}\ \bibnamefont
  {Cumming}}} (\bibinfo {year} {2015}),\ \bibfield  {title} {\enquote {\bibinfo
  {title} {{Time-dependent, compositionally driven convection in the oceans of
  accreting neutron stars}},}\ }\href {\doibase 10.1088/0004-637X/802/1/29}
  {\bibfield  {journal} {\bibinfo  {journal} {Astrophys. J.}\ }\textbf
  {\bibinfo {volume} {802}}~(\bibinfo {number} {1}),\ \bibinfo {pages} {29}},\
  \Eprint {http://arxiv.org/abs/1409.2533} {arXiv:1409.2533 [astro-ph.SR]}
  \BibitemShut {NoStop}%
\bibitem [{\citenamefont {Merritt}\ \emph {et~al.}(2016)\citenamefont
  {Merritt}, \citenamefont {Cackett}, \citenamefont {Brown}, \citenamefont
  {Page}, \citenamefont {Cumming}, \citenamefont {Degenaar}, \citenamefont
  {Deibel}, \citenamefont {Homan}, \citenamefont {Miller},\ and\ \citenamefont
  {Wijnands}}]{0004-637X-833-2-186}%
  \BibitemOpen
  \bibfield  {author} {\bibinfo {author} {\bibnamefont {Merritt}, \bibfnamefont
  {Rachael~L}}, \bibinfo {author} {\bibfnamefont {Edward~M.}\ \bibnamefont
  {Cackett}}, \bibinfo {author} {\bibfnamefont {Edward~F.}\ \bibnamefont
  {Brown}}, \bibinfo {author} {\bibfnamefont {Dany}\ \bibnamefont {Page}},
  \bibinfo {author} {\bibfnamefont {Andrew}\ \bibnamefont {Cumming}}, \bibinfo
  {author} {\bibfnamefont {Nathalie}\ \bibnamefont {Degenaar}}, \bibinfo
  {author} {\bibfnamefont {Alex}\ \bibnamefont {Deibel}}, \bibinfo {author}
  {\bibfnamefont {Jeroen}\ \bibnamefont {Homan}}, \bibinfo {author}
  {\bibfnamefont {Jon~M.}\ \bibnamefont {Miller}}, \ and\ \bibinfo {author}
  {\bibfnamefont {Rudy}\ \bibnamefont {Wijnands}}} (\bibinfo {year} {2016}),\
  \bibfield  {title} {\enquote {\bibinfo {title} {The thermal state of ks
  1731−260 after 14.5 years in quiescence},}\ }\href
  {http://stacks.iop.org/0004-637X/833/i=2/a=186} {\bibfield  {journal}
  {\bibinfo  {journal} {The Astrophysical Journal}\ }\textbf {\bibinfo {volume}
  {833}}~(\bibinfo {number} {2}),\ \bibinfo {pages} {186}}\BibitemShut
  {NoStop}%
\bibitem [{\citenamefont {Metcalfe}(2005)}]{astroseismology}%
  \BibitemOpen
  \bibfield  {author} {\bibinfo {author} {\bibnamefont {Metcalfe},
  \bibfnamefont {T~S}}} (\bibinfo {year} {2005}),\ \href@noop {} {\bibfield
  {journal} {\bibinfo  {journal} {Monthly Notice Royal Astron. Soc.}\ }\textbf
  {\bibinfo {volume} {363}},\ \bibinfo {pages} {L86}}\BibitemShut {NoStop}%
\bibitem [{\citenamefont {Nakazato}\ \emph {et~al.}(2009)\citenamefont
  {Nakazato}, \citenamefont {Oyamatsu},\ and\ \citenamefont
  {Yamada}}]{PhysRevLett.103.132501}%
  \BibitemOpen
  \bibfield  {author} {\bibinfo {author} {\bibnamefont {Nakazato},
  \bibfnamefont {Kenichiro}}, \bibinfo {author} {\bibfnamefont {Kazuhiro}\
  \bibnamefont {Oyamatsu}}, \ and\ \bibinfo {author} {\bibfnamefont {Shoichi}\
  \bibnamefont {Yamada}}} (\bibinfo {year} {2009}),\ \bibfield  {title}
  {\enquote {\bibinfo {title} {Gyroid phase in nuclear pasta},}\ }\href
  {\doibase 10.1103/PhysRevLett.103.132501} {\bibfield  {journal} {\bibinfo
  {journal} {Phys. Rev. Lett.}\ }\textbf {\bibinfo {volume} {103}},\ \bibinfo
  {pages} {132501}}\BibitemShut {NoStop}%
\bibitem [{\citenamefont {Newton}\ and\ \citenamefont
  {Stone}(2009)}]{PhysRevC.79.055801}%
  \BibitemOpen
  \bibfield  {author} {\bibinfo {author} {\bibnamefont {Newton}, \bibfnamefont
  {W~G}}, \ and\ \bibinfo {author} {\bibfnamefont {J.~R.}\ \bibnamefont
  {Stone}}} (\bibinfo {year} {2009}),\ \bibfield  {title} {\enquote {\bibinfo
  {title} {Modeling nuclear ``pasta'' and the transition to uniform nuclear
  matter with the 3d skyrme-hartree-fock method at finite temperature:
  Core-collapse supernovae},}\ }\href {\doibase 10.1103/PhysRevC.79.055801}
  {\bibfield  {journal} {\bibinfo  {journal} {Phys. Rev. C}\ }\textbf {\bibinfo
  {volume} {79}},\ \bibinfo {pages} {055801}}\BibitemShut {NoStop}%
\bibitem [{\citenamefont {Oyamatsu}(1993)}]{OYAMATSU1993431}%
  \BibitemOpen
  \bibfield  {author} {\bibinfo {author} {\bibnamefont {Oyamatsu},
  \bibfnamefont {K}}} (\bibinfo {year} {1993}),\ \bibfield  {title} {\enquote
  {\bibinfo {title} {Nuclear shapes in the inner crust of a neutron star},}\
  }\href {\doibase http://dx.doi.org/10.1016/0375-9474(93)90020-X} {\bibfield
  {journal} {\bibinfo  {journal} {Nuclear Physics A}\ }\textbf {\bibinfo
  {volume} {561}}~(\bibinfo {number} {3}),\ \bibinfo {pages} {431 --
  452}}\BibitemShut {NoStop}%
\bibitem [{\citenamefont {Patruno}\ \emph {et~al.}(2017)\citenamefont
  {Patruno}, \citenamefont {Haskell},\ and\ \citenamefont
  {Andersson}}]{Patruno:2017oum}%
  \BibitemOpen
  \bibfield  {author} {\bibinfo {author} {\bibnamefont {Patruno}, \bibfnamefont
  {A}}, \bibinfo {author} {\bibfnamefont {B.}~\bibnamefont {Haskell}}, \ and\
  \bibinfo {author} {\bibfnamefont {N.}~\bibnamefont {Andersson}}} (\bibinfo
  {year} {2017}),\ \bibfield  {title} {\enquote {\bibinfo {title} {{The Spin
  Distribution of Fast Spinning Neutron Stars in Low Mass X-Ray Binaries:
  Evidence for Two Sub-Populations}},}\ }\href@noop {} {\ }\Eprint
  {http://arxiv.org/abs/1705.07669} {arXiv:1705.07669 [astro-ph.HE]}
  \BibitemShut {NoStop}%
\bibitem [{\citenamefont {Pethick}\ and\ \citenamefont
  {Potekhin}(1998)}]{Pethick19987}%
  \BibitemOpen
  \bibfield  {author} {\bibinfo {author} {\bibnamefont {Pethick}, \bibfnamefont
  {C~J}}, \ and\ \bibinfo {author} {\bibfnamefont {A.~Y.}\ \bibnamefont
  {Potekhin}}} (\bibinfo {year} {1998}),\ \bibfield  {title} {\enquote
  {\bibinfo {title} {Liquid crystals in the mantles of neutron stars},}\ }\href
  {\doibase http://dx.doi.org/10.1016/S0370-2693(98)00341-4} {\bibfield
  {journal} {\bibinfo  {journal} {Physics Letters B}\ }\textbf {\bibinfo
  {volume} {427}}~(\bibinfo {number} {1–2}),\ \bibinfo {pages} {7 --
  12}}\BibitemShut {NoStop}%
\bibitem [{\citenamefont {Pethick}\ and\ \citenamefont
  {Ravenhall}(1995)}]{pethick1995}%
  \BibitemOpen
  \bibfield  {author} {\bibinfo {author} {\bibnamefont {Pethick}, \bibfnamefont
  {C~J}}, \ and\ \bibinfo {author} {\bibfnamefont {D~G}\ \bibnamefont
  {Ravenhall}}} (\bibinfo {year} {1995}),\ \bibfield  {title} {\enquote
  {\bibinfo {title} {Matter at large neutron excess and the physics of
  neutron-star crusts},}\ }\href {\doibase 10.1146/annurev.ns.45.120195.002241}
  {\bibfield  {journal} {\bibinfo  {journal} {Annual Review of Nuclear and
  Particle Science}\ }\textbf {\bibinfo {volume} {45}}~(\bibinfo {number}
  {1}),\ \bibinfo {pages} {429--484}},\ \Eprint
  {http://arxiv.org/abs/http://dx.doi.org/10.1146/annurev.ns.45.120195.002241}
  {http://dx.doi.org/10.1146/annurev.ns.45.120195.002241} \BibitemShut
  {NoStop}%
\bibitem [{\citenamefont {Pollock}\ and\ \citenamefont
  {Hansen}(1973)}]{Pollock73}%
  \BibitemOpen
  \bibfield  {author} {\bibinfo {author} {\bibnamefont {Pollock}, \bibfnamefont
  {E~L}}, \ and\ \bibinfo {author} {\bibfnamefont {J.~P.}\ \bibnamefont
  {Hansen}}} (\bibinfo {year} {1973}),\ \bibfield  {title} {\enquote {\bibinfo
  {title} {Statistical mechanics of dense ionized matter. ii. equilibrium
  properties and melting transition of the crystallized one-component
  plasma},}\ }\href {\doibase 10.1103/PhysRevA.8.3110} {\bibfield  {journal}
  {\bibinfo  {journal} {Phys. Rev. A}\ }\textbf {\bibinfo {volume} {8}},\
  \bibinfo {pages} {3110--3122}}\BibitemShut {NoStop}%
\bibitem [{\citenamefont {{Pons}}\ \emph {et~al.}(2013)\citenamefont {{Pons}},
  \citenamefont {{Vigan{\`o}}},\ and\ \citenamefont
  {{Rea}}}]{2013NatPh...9..431P}%
  \BibitemOpen
  \bibfield  {author} {\bibinfo {author} {\bibnamefont {{Pons}}, \bibfnamefont
  {J~A}}, \bibinfo {author} {\bibfnamefont {D.}~\bibnamefont {{Vigan{\`o}}}}, \
  and\ \bibinfo {author} {\bibfnamefont {N.}~\bibnamefont {{Rea}}}} (\bibinfo
  {year} {2013}),\ \bibfield  {title} {\enquote {\bibinfo {title} {{A highly
  resistive layer within the crust of X-ray pulsars limits their spin
  periods}},}\ }\href {\doibase 10.1038/nphys2640} {\bibfield  {journal}
  {\bibinfo  {journal} {Nature Physics}\ }\textbf {\bibinfo {volume} {9}},\
  \bibinfo {pages} {431--434}},\ \Eprint {http://arxiv.org/abs/1304.6546}
  {arXiv:1304.6546 [astro-ph.SR]} \BibitemShut {NoStop}%
\bibitem [{\citenamefont {Ravenhall}\ \emph {et~al.}(1983)\citenamefont
  {Ravenhall}, \citenamefont {Pethick},\ and\ \citenamefont
  {Wilson}}]{PhysRevLett.50.2066}%
  \BibitemOpen
  \bibfield  {author} {\bibinfo {author} {\bibnamefont {Ravenhall},
  \bibfnamefont {D~G}}, \bibinfo {author} {\bibfnamefont {C.~J.}\ \bibnamefont
  {Pethick}}, \ and\ \bibinfo {author} {\bibfnamefont {J.~R.}\ \bibnamefont
  {Wilson}}} (\bibinfo {year} {1983}),\ \bibfield  {title} {\enquote {\bibinfo
  {title} {Structure of matter below nuclear saturation density},}\ }\href
  {\doibase 10.1103/PhysRevLett.50.2066} {\bibfield  {journal} {\bibinfo
  {journal} {Phys. Rev. Lett.}\ }\textbf {\bibinfo {volume} {50}},\ \bibinfo
  {pages} {2066--2069}}\BibitemShut {NoStop}%
\bibitem [{\citenamefont {Renedo}\ \emph {et~al.}(2010)\citenamefont {Renedo},
  \citenamefont {Althaus}, \citenamefont {Bertolami}, \citenamefont {Romero},
  \citenamefont {Co—rsico}, \citenamefont {Rohrmann},\ and\ \citenamefont
  {Garci'a-Berro}}]{renedo}%
  \BibitemOpen
  \bibfield  {author} {\bibinfo {author} {\bibnamefont {Renedo}, \bibfnamefont
  {I}}, \bibinfo {author} {\bibfnamefont {L.~G.}\ \bibnamefont {Althaus}},
  \bibinfo {author} {\bibfnamefont {M.~M.~Miller}\ \bibnamefont {Bertolami}},
  \bibinfo {author} {\bibfnamefont {A.~D.}\ \bibnamefont {Romero}}, \bibinfo
  {author} {\bibfnamefont {A.~H.}\ \bibnamefont {Co—rsico}}, \bibinfo {author}
  {\bibfnamefont {R.~D.}\ \bibnamefont {Rohrmann}}, \ and\ \bibinfo {author}
  {\bibfnamefont {E.}~\bibnamefont {Garci'a-Berro}}} (\bibinfo {year} {2010}),\
  \href@noop {} {\bibfield  {journal} {\bibinfo  {journal} {Astrophys. J.}\
  }\textbf {\bibinfo {volume} {717}},\ \bibinfo {pages} {183}}\BibitemShut
  {NoStop}%
\bibitem [{\citenamefont {Rutledge}\ and\ \citenamefont
  {et~al.}(2002)}]{crustcooling2}%
  \BibitemOpen
  \bibfield  {author} {\bibinfo {author} {\bibnamefont {Rutledge},
  \bibfnamefont {R~E}}, \ and\ \bibinfo {author} {\bibnamefont {et~al.}}}
  (\bibinfo {year} {2002}),\ \href@noop {} {\bibfield  {journal} {\bibinfo
  {journal} {ApJ.}\ }\textbf {\bibinfo {volume} {580}},\ \bibinfo {pages}
  {413}}\BibitemShut {NoStop}%
\bibitem [{\citenamefont {Sagert}\ \emph {et~al.}(2016)\citenamefont {Sagert},
  \citenamefont {Fann}, \citenamefont {Fattoyev}, \citenamefont {Postnikov},\
  and\ \citenamefont {Horowitz}}]{PhysRevC.93.055801}%
  \BibitemOpen
  \bibfield  {author} {\bibinfo {author} {\bibnamefont {Sagert}, \bibfnamefont
  {I}}, \bibinfo {author} {\bibfnamefont {G.~I.}\ \bibnamefont {Fann}},
  \bibinfo {author} {\bibfnamefont {F.~J.}\ \bibnamefont {Fattoyev}}, \bibinfo
  {author} {\bibfnamefont {S.}~\bibnamefont {Postnikov}}, \ and\ \bibinfo
  {author} {\bibfnamefont {C.~J.}\ \bibnamefont {Horowitz}}} (\bibinfo {year}
  {2016}),\ \bibfield  {title} {\enquote {\bibinfo {title} {Quantum simulations
  of nuclei and nuclear pasta with the multiresolution adaptive numerical
  environment for scientific simulations},}\ }\href {\doibase
  10.1103/PhysRevC.93.055801} {\bibfield  {journal} {\bibinfo  {journal} {Phys.
  Rev. C}\ }\textbf {\bibinfo {volume} {93}},\ \bibinfo {pages}
  {055801}}\BibitemShut {NoStop}%
\bibitem [{\citenamefont {Salaris}\ \emph {et~al.}(2010)\citenamefont
  {Salaris}, \citenamefont {Cassisi}, \citenamefont {Pietrinferni},
  \citenamefont {Kowalski},\ and\ \citenamefont {Isern}}]{salaris1}%
  \BibitemOpen
  \bibfield  {author} {\bibinfo {author} {\bibnamefont {Salaris}, \bibfnamefont
  {M}}, \bibinfo {author} {\bibfnamefont {S.}~\bibnamefont {Cassisi}}, \bibinfo
  {author} {\bibfnamefont {A.}~\bibnamefont {Pietrinferni}}, \bibinfo {author}
  {\bibfnamefont {P.~M.}\ \bibnamefont {Kowalski}}, \ and\ \bibinfo {author}
  {\bibfnamefont {J.}~\bibnamefont {Isern}}} (\bibinfo {year} {2010}),\
  \href@noop {} {\bibfield  {journal} {\bibinfo  {journal} {Astrophys. J.}\
  }\textbf {\bibinfo {volume} {716}},\ \bibinfo {pages} {1241}}\BibitemShut
  {NoStop}%
\bibitem [{\citenamefont {Salaris}\ \emph {et~al.}(1997)\citenamefont
  {Salaris}, \citenamefont {Dom'ingues}, \citenamefont {Garci'a-Berro},
  \citenamefont {Hernanz}, \citenamefont {Isern},\ and\ \citenamefont
  {Mochkovitch}}]{salaris}%
  \BibitemOpen
  \bibfield  {author} {\bibinfo {author} {\bibnamefont {Salaris}, \bibfnamefont
  {Maurizio}}, \bibinfo {author} {\bibfnamefont {Inmaculada}\ \bibnamefont
  {Dom'ingues}}, \bibinfo {author} {\bibfnamefont {Enrique}\ \bibnamefont
  {Garci'a-Berro}}, \bibinfo {author} {\bibfnamefont {Margarida}\ \bibnamefont
  {Hernanz}}, \bibinfo {author} {\bibfnamefont {Jordi}\ \bibnamefont {Isern}},
  \ and\ \bibinfo {author} {\bibfnamefont {Robert}\ \bibnamefont
  {Mochkovitch}}} (\bibinfo {year} {1997}),\ \href@noop {} {\bibfield
  {journal} {\bibinfo  {journal} {Astrophys. J}\ }\textbf {\bibinfo {volume}
  {486}},\ \bibinfo {pages} {413}}\BibitemShut {NoStop}%
\bibitem [{\citenamefont {Schatz}\ and\ \citenamefont {et~al}(2001)}]{rpash}%
  \BibitemOpen
  \bibfield  {author} {\bibinfo {author} {\bibnamefont {Schatz}, \bibfnamefont
  {H}}, \ and\ \bibinfo {author} {\bibnamefont {et~al}}} (\bibinfo {year}
  {2001}),\ \href@noop {} {\bibfield  {journal} {\bibinfo  {journal} {Phys.
  Rev. Lett.}\ }\textbf {\bibinfo {volume} {86}},\ \bibinfo {pages}
  {3471}}\BibitemShut {NoStop}%
\bibitem [{\citenamefont {Schneider}\ \emph {et~al.}(2014)\citenamefont
  {Schneider}, \citenamefont {Berry}, \citenamefont {Briggs}, \citenamefont
  {Caplan},\ and\ \citenamefont {Horowitz}}]{PhysRevC.90.055805}%
  \BibitemOpen
  \bibfield  {author} {\bibinfo {author} {\bibnamefont {Schneider},
  \bibfnamefont {A~S}}, \bibinfo {author} {\bibfnamefont {D.~K.}\ \bibnamefont
  {Berry}}, \bibinfo {author} {\bibfnamefont {C.~M.}\ \bibnamefont {Briggs}},
  \bibinfo {author} {\bibfnamefont {M.~E.}\ \bibnamefont {Caplan}}, \ and\
  \bibinfo {author} {\bibfnamefont {C.~J.}\ \bibnamefont {Horowitz}}} (\bibinfo
  {year} {2014}),\ \bibfield  {title} {\enquote {\bibinfo {title} {Nuclear
  ``waffles''},}\ }\href {\doibase 10.1103/PhysRevC.90.055805} {\bibfield
  {journal} {\bibinfo  {journal} {Phys. Rev. C}\ }\textbf {\bibinfo {volume}
  {90}},\ \bibinfo {pages} {055805}}\BibitemShut {NoStop}%
\bibitem [{\citenamefont {Schneider}\ \emph {et~al.}(2016)\citenamefont
  {Schneider}, \citenamefont {Berry}, \citenamefont {Caplan}, \citenamefont
  {Horowitz},\ and\ \citenamefont {Lin}}]{PhysRevC.93.065806}%
  \BibitemOpen
  \bibfield  {author} {\bibinfo {author} {\bibnamefont {Schneider},
  \bibfnamefont {A~S}}, \bibinfo {author} {\bibfnamefont {D.~K.}\ \bibnamefont
  {Berry}}, \bibinfo {author} {\bibfnamefont {M.~E.}\ \bibnamefont {Caplan}},
  \bibinfo {author} {\bibfnamefont {C.~J.}\ \bibnamefont {Horowitz}}, \ and\
  \bibinfo {author} {\bibfnamefont {Z.}~\bibnamefont {Lin}}} (\bibinfo {year}
  {2016}),\ \bibfield  {title} {\enquote {\bibinfo {title} {Effect of
  topological defects on ``nuclear pasta'' observables},}\ }\href {\doibase
  10.1103/PhysRevC.93.065806} {\bibfield  {journal} {\bibinfo  {journal} {Phys.
  Rev. C}\ }\textbf {\bibinfo {volume} {93}},\ \bibinfo {pages}
  {065806}}\BibitemShut {NoStop}%
\bibitem [{\citenamefont {Schneider}\ \emph {et~al.}(2013)\citenamefont
  {Schneider}, \citenamefont {Horowitz}, \citenamefont {Hughto},\ and\
  \citenamefont {Berry}}]{PhysRevC.88.065807}%
  \BibitemOpen
  \bibfield  {author} {\bibinfo {author} {\bibnamefont {Schneider},
  \bibfnamefont {A~S}}, \bibinfo {author} {\bibfnamefont {C.~J.}\ \bibnamefont
  {Horowitz}}, \bibinfo {author} {\bibfnamefont {J.}~\bibnamefont {Hughto}}, \
  and\ \bibinfo {author} {\bibfnamefont {D.~K.}\ \bibnamefont {Berry}}}
  (\bibinfo {year} {2013}),\ \bibfield  {title} {\enquote {\bibinfo {title}
  {Nuclear ``pasta'' formation},}\ }\href {\doibase 10.1103/PhysRevC.88.065807}
  {\bibfield  {journal} {\bibinfo  {journal} {Phys. Rev. C}\ }\textbf {\bibinfo
  {volume} {88}},\ \bibinfo {pages} {065807}}\BibitemShut {NoStop}%
\bibitem [{\citenamefont {Scholberg}(2012)}]{Scholberg2012}%
  \BibitemOpen
  \bibfield  {author} {\bibinfo {author} {\bibnamefont {Scholberg},
  \bibfnamefont {Kate}}} (\bibinfo {year} {2012}),\ \bibfield  {title}
  {\enquote {\bibinfo {title} {Supernova neutrino detection},}\ }\href@noop {}
  {\bibfield  {journal} {\bibinfo  {journal} {Ann. Rev. Nuclear and Particle
  Science}\ }\textbf {\bibinfo {volume} {62}},\ \bibinfo {pages}
  {81}}\BibitemShut {NoStop}%
\bibitem [{\citenamefont {Schuetrumpf}\ \emph {et~al.}(2013)\citenamefont
  {Schuetrumpf}, \citenamefont {Klatt}, \citenamefont {Iida}, \citenamefont
  {Maruhn}, \citenamefont {Mecke},\ and\ \citenamefont
  {Reinhard}}]{PhysRevC.87.055805}%
  \BibitemOpen
  \bibfield  {author} {\bibinfo {author} {\bibnamefont {Schuetrumpf},
  \bibfnamefont {B}}, \bibinfo {author} {\bibfnamefont {M.~A.}\ \bibnamefont
  {Klatt}}, \bibinfo {author} {\bibfnamefont {K.}~\bibnamefont {Iida}},
  \bibinfo {author} {\bibfnamefont {J.~A.}\ \bibnamefont {Maruhn}}, \bibinfo
  {author} {\bibfnamefont {K.}~\bibnamefont {Mecke}}, \ and\ \bibinfo {author}
  {\bibfnamefont {P.-G.}\ \bibnamefont {Reinhard}}} (\bibinfo {year} {2013}),\
  \bibfield  {title} {\enquote {\bibinfo {title} {Time-dependent hartree-fock
  approach to nuclear ``pasta'' at finite temperature},}\ }\href {\doibase
  10.1103/PhysRevC.87.055805} {\bibfield  {journal} {\bibinfo  {journal} {Phys.
  Rev. C}\ }\textbf {\bibinfo {volume} {87}},\ \bibinfo {pages}
  {055805}}\BibitemShut {NoStop}%
\bibitem [{\citenamefont {Schuetrumpf}\ \emph {et~al.}(2015)\citenamefont
  {Schuetrumpf}, \citenamefont {Klatt}, \citenamefont {Iida}, \citenamefont
  {Schr\"oder-Turk}, \citenamefont {Maruhn}, \citenamefont {Mecke},\ and\
  \citenamefont {Reinhard}}]{PhysRevC.91.025801}%
  \BibitemOpen
  \bibfield  {author} {\bibinfo {author} {\bibnamefont {Schuetrumpf},
  \bibfnamefont {B}}, \bibinfo {author} {\bibfnamefont {M.~A.}\ \bibnamefont
  {Klatt}}, \bibinfo {author} {\bibfnamefont {K.}~\bibnamefont {Iida}},
  \bibinfo {author} {\bibfnamefont {G.~E.}\ \bibnamefont {Schr\"oder-Turk}},
  \bibinfo {author} {\bibfnamefont {J.~A.}\ \bibnamefont {Maruhn}}, \bibinfo
  {author} {\bibfnamefont {K.}~\bibnamefont {Mecke}}, \ and\ \bibinfo {author}
  {\bibfnamefont {P.-G.}\ \bibnamefont {Reinhard}}} (\bibinfo {year} {2015}),\
  \bibfield  {title} {\enquote {\bibinfo {title} {Appearance of the single
  gyroid network phase in ``nuclear pasta'' matter},}\ }\href {\doibase
  10.1103/PhysRevC.91.025801} {\bibfield  {journal} {\bibinfo  {journal} {Phys.
  Rev. C}\ }\textbf {\bibinfo {volume} {91}},\ \bibinfo {pages}
  {025801}}\BibitemShut {NoStop}%
\bibitem [{\citenamefont {Schuetrumpf}\ and\ \citenamefont
  {Nazarewicz}(2015)}]{PhysRevC.92.045806}%
  \BibitemOpen
  \bibfield  {author} {\bibinfo {author} {\bibnamefont {Schuetrumpf},
  \bibfnamefont {B}}, \ and\ \bibinfo {author} {\bibfnamefont {W.}~\bibnamefont
  {Nazarewicz}}} (\bibinfo {year} {2015}),\ \bibfield  {title} {\enquote
  {\bibinfo {title} {Twist-averaged boundary conditions for nuclear pasta
  hartree-fock calculations},}\ }\href {\doibase 10.1103/PhysRevC.92.045806}
  {\bibfield  {journal} {\bibinfo  {journal} {Phys. Rev. C}\ }\textbf {\bibinfo
  {volume} {92}},\ \bibinfo {pages} {045806}}\BibitemShut {NoStop}%
\bibitem [{\citenamefont {Seddon}(1990)}]{SEDDON19901}%
  \BibitemOpen
  \bibfield  {author} {\bibinfo {author} {\bibnamefont {Seddon}, \bibfnamefont
  {John~M}}} (\bibinfo {year} {1990}),\ \bibfield  {title} {\enquote {\bibinfo
  {title} {Structure of the inverted hexagonal (hii) phase, and non-lamellar
  phase transitions of lipids},}\ }\href {\doibase
  http://dx.doi.org/10.1016/0304-4157(90)90002-T} {\bibfield  {journal}
  {\bibinfo  {journal} {Biochimica et Biophysica Acta (BBA) - Reviews on
  Biomembranes}\ }\textbf {\bibinfo {volume} {1031}}~(\bibinfo {number} {1}),\
  \bibinfo {pages} {1 -- 69}}\BibitemShut {NoStop}%
\bibitem [{\citenamefont {Shternin}\ \emph {et~al.}(2007)\citenamefont
  {Shternin}, \citenamefont {Yakovlev}, \citenamefont {Haensel},\ and\
  \citenamefont {Potekhin}}]{crustcooling3}%
  \BibitemOpen
  \bibfield  {author} {\bibinfo {author} {\bibnamefont {Shternin},
  \bibfnamefont {P~S}}, \bibinfo {author} {\bibfnamefont {D.~G.}\ \bibnamefont
  {Yakovlev}}, \bibinfo {author} {\bibfnamefont {P.}~\bibnamefont {Haensel}}, \
  and\ \bibinfo {author} {\bibfnamefont {A.~Y.}\ \bibnamefont {Potekhin}}}
  (\bibinfo {year} {2007}),\ \href@noop {} {\bibfield  {journal} {\bibinfo
  {journal} {Mon. Not. Roy. Astron. Soc.}\ }\textbf {\bibinfo {volume} {382}},\
  \bibinfo {pages} {L43}}\BibitemShut {NoStop}%
\bibitem [{\citenamefont {Slattery}\ \emph {et~al.}(1980)\citenamefont
  {Slattery}, \citenamefont {Doolen},\ and\ \citenamefont
  {DeWitt}}]{Slattery80}%
  \BibitemOpen
  \bibfield  {author} {\bibinfo {author} {\bibnamefont {Slattery},
  \bibfnamefont {W~L}}, \bibinfo {author} {\bibfnamefont {G.~D.}\ \bibnamefont
  {Doolen}}, \ and\ \bibinfo {author} {\bibfnamefont {H.~E.}\ \bibnamefont
  {DeWitt}}} (\bibinfo {year} {1980}),\ \bibfield  {title} {\enquote {\bibinfo
  {title} {Improved equation of state for the classical one-component
  plasma},}\ }\href {\doibase 10.1103/PhysRevA.21.2087} {\bibfield  {journal}
  {\bibinfo  {journal} {Phys. Rev. A}\ }\textbf {\bibinfo {volume} {21}},\
  \bibinfo {pages} {2087--2095}}\BibitemShut {NoStop}%
\bibitem [{\citenamefont {Sonoda}\ \emph {et~al.}(2008)\citenamefont {Sonoda},
  \citenamefont {Watanabe}, \citenamefont {Sato}, \citenamefont {Yasuoka},\
  and\ \citenamefont {Ebisuzaki}}]{PhysRevC.77.035806}%
  \BibitemOpen
  \bibfield  {author} {\bibinfo {author} {\bibnamefont {Sonoda}, \bibfnamefont
  {Hidetaka}}, \bibinfo {author} {\bibfnamefont {Gentaro}\ \bibnamefont
  {Watanabe}}, \bibinfo {author} {\bibfnamefont {Katsuhiko}\ \bibnamefont
  {Sato}}, \bibinfo {author} {\bibfnamefont {Kenji}\ \bibnamefont {Yasuoka}}, \
  and\ \bibinfo {author} {\bibfnamefont {Toshikazu}\ \bibnamefont {Ebisuzaki}}}
  (\bibinfo {year} {2008}),\ \bibfield  {title} {\enquote {\bibinfo {title}
  {Phase diagram of nuclear ``pasta'' and its uncertainties in supernova
  cores},}\ }\href {\doibase 10.1103/PhysRevC.77.035806} {\bibfield  {journal}
  {\bibinfo  {journal} {Phys. Rev. C}\ }\textbf {\bibinfo {volume} {77}},\
  \bibinfo {pages} {035806}}\BibitemShut {NoStop}%
\bibitem [{\citenamefont {Sun}\ \emph {et~al.}(2016)\citenamefont {Sun},
  \citenamefont {Melatos}, \citenamefont {Lasky}, \citenamefont {Chung},\ and\
  \citenamefont {Darman}}]{PhysRevD.94.082004}%
  \BibitemOpen
  \bibfield  {author} {\bibinfo {author} {\bibnamefont {Sun}, \bibfnamefont
  {L}}, \bibinfo {author} {\bibfnamefont {A.}~\bibnamefont {Melatos}}, \bibinfo
  {author} {\bibfnamefont {P.~D.}\ \bibnamefont {Lasky}}, \bibinfo {author}
  {\bibfnamefont {C.~T.~Y.}\ \bibnamefont {Chung}}, \ and\ \bibinfo {author}
  {\bibfnamefont {N.~S.}\ \bibnamefont {Darman}}} (\bibinfo {year} {2016}),\
  \bibfield  {title} {\enquote {\bibinfo {title} {Cross-correlation search for
  continuous gravitational waves from a compact object in snr 1987a in ligo
  science run 5},}\ }\href {\doibase 10.1103/PhysRevD.94.082004} {\bibfield
  {journal} {\bibinfo  {journal} {Phys. Rev. D}\ }\textbf {\bibinfo {volume}
  {94}},\ \bibinfo {pages} {082004}}\BibitemShut {NoStop}%
\bibitem [{\citenamefont {Terasaki}\ \emph {et~al.}(2013)\citenamefont
  {Terasaki}, \citenamefont {Shemesh}, \citenamefont {Kasthuri}, \citenamefont
  {Klemm}, \citenamefont {Schalek}, \citenamefont {Hayworth}, \citenamefont
  {Hand}, \citenamefont {Yankova}, \citenamefont {Huber}, \citenamefont
  {Lichtman}, \citenamefont {Rapoport},\ and\ \citenamefont
  {Kozlov}}]{Terasaki2013285}%
  \BibitemOpen
  \bibfield  {author} {\bibinfo {author} {\bibnamefont {Terasaki},
  \bibfnamefont {Mark}}, \bibinfo {author} {\bibfnamefont {Tom}\ \bibnamefont
  {Shemesh}}, \bibinfo {author} {\bibfnamefont {Narayanan}\ \bibnamefont
  {Kasthuri}}, \bibinfo {author} {\bibfnamefont {Robin~W.}\ \bibnamefont
  {Klemm}}, \bibinfo {author} {\bibfnamefont {Richard}\ \bibnamefont
  {Schalek}}, \bibinfo {author} {\bibfnamefont {Kenneth~J.}\ \bibnamefont
  {Hayworth}}, \bibinfo {author} {\bibfnamefont {Arthur~R.}\ \bibnamefont
  {Hand}}, \bibinfo {author} {\bibfnamefont {Maya}\ \bibnamefont {Yankova}},
  \bibinfo {author} {\bibfnamefont {Greg}\ \bibnamefont {Huber}}, \bibinfo
  {author} {\bibfnamefont {Jeff~W.}\ \bibnamefont {Lichtman}}, \bibinfo
  {author} {\bibfnamefont {Tom~A.}\ \bibnamefont {Rapoport}}, \ and\ \bibinfo
  {author} {\bibfnamefont {Michael~M.}\ \bibnamefont {Kozlov}}} (\bibinfo
  {year} {2013}),\ \bibfield  {title} {\enquote {\bibinfo {title} {Stacked
  endoplasmic reticulum sheets are connected by helicoidal membrane motifs},}\
  }\href {\doibase http://dx.doi.org/10.1016/j.cell.2013.06.031} {\bibfield
  {journal} {\bibinfo  {journal} {Cell}\ }\textbf {\bibinfo {volume}
  {154}}~(\bibinfo {number} {2}),\ \bibinfo {pages} {285 -- 296}}\BibitemShut
  {NoStop}%
\bibitem [{\citenamefont {Verlet}(1967)}]{PhysRev.159.98}%
  \BibitemOpen
  \bibfield  {author} {\bibinfo {author} {\bibnamefont {Verlet}, \bibfnamefont
  {Loup}}} (\bibinfo {year} {1967}),\ \bibfield  {title} {\enquote {\bibinfo
  {title} {Computer "experiments" on classical fluids. i. thermodynamical
  properties of lennard-jones molecules},}\ }\href {\doibase
  10.1103/PhysRev.159.98} {\bibfield  {journal} {\bibinfo  {journal} {Phys.
  Rev.}\ }\textbf {\bibinfo {volume} {159}},\ \bibinfo {pages}
  {98--103}}\BibitemShut {NoStop}%
\bibitem [{\citenamefont {Vicentini}\ \emph {et~al.}(1985)\citenamefont
  {Vicentini}, \citenamefont {Jacucci},\ and\ \citenamefont
  {Pandharipande}}]{PhysRevC.31.1783}%
  \BibitemOpen
  \bibfield  {author} {\bibinfo {author} {\bibnamefont {Vicentini},
  \bibfnamefont {A}}, \bibinfo {author} {\bibfnamefont {G.}~\bibnamefont
  {Jacucci}}, \ and\ \bibinfo {author} {\bibfnamefont {V.~R.}\ \bibnamefont
  {Pandharipande}}} (\bibinfo {year} {1985}),\ \bibfield  {title} {\enquote
  {\bibinfo {title} {Fragmentation of hot classical drops},}\ }\href {\doibase
  10.1103/PhysRevC.31.1783} {\bibfield  {journal} {\bibinfo  {journal} {Phys.
  Rev. C}\ }\textbf {\bibinfo {volume} {31}},\ \bibinfo {pages}
  {1783--1793}}\BibitemShut {NoStop}%
\bibitem [{\citenamefont {Watanabe}\ \emph {et~al.}(2000)\citenamefont
  {Watanabe}, \citenamefont {Iida},\ and\ \citenamefont
  {Sato}}]{WATANABE2000455}%
  \BibitemOpen
  \bibfield  {author} {\bibinfo {author} {\bibnamefont {Watanabe},
  \bibfnamefont {Gentaro}}, \bibinfo {author} {\bibfnamefont {Kei}\
  \bibnamefont {Iida}}, \ and\ \bibinfo {author} {\bibfnamefont {Katsuhiko}\
  \bibnamefont {Sato}}} (\bibinfo {year} {2000}),\ \bibfield  {title} {\enquote
  {\bibinfo {title} {Thermodynamic properties of nuclear €œpasta€ in neutron
  star crusts},}\ }\href {\doibase
  http://dx.doi.org/10.1016/S0375-9474(00)00197-4} {\bibfield  {journal}
  {\bibinfo  {journal} {Nuclear Physics A}\ }\textbf {\bibinfo {volume}
  {676}}~(\bibinfo {number} {1}),\ \bibinfo {pages} {455 -- 473}}\BibitemShut
  {NoStop}%
\bibitem [{\citenamefont {Watanabe}\ \emph {et~al.}(2005)\citenamefont
  {Watanabe}, \citenamefont {Maruyama}, \citenamefont {Sato}, \citenamefont
  {Yasuoka},\ and\ \citenamefont {Ebisuzaki}}]{PhysRevLett.94.031101}%
  \BibitemOpen
  \bibfield  {author} {\bibinfo {author} {\bibnamefont {Watanabe},
  \bibfnamefont {Gentaro}}, \bibinfo {author} {\bibfnamefont {Toshiki}\
  \bibnamefont {Maruyama}}, \bibinfo {author} {\bibfnamefont {Katsuhiko}\
  \bibnamefont {Sato}}, \bibinfo {author} {\bibfnamefont {Kenji}\ \bibnamefont
  {Yasuoka}}, \ and\ \bibinfo {author} {\bibfnamefont {Toshikazu}\ \bibnamefont
  {Ebisuzaki}}} (\bibinfo {year} {2005}),\ \bibfield  {title} {\enquote
  {\bibinfo {title} {Simulation of transitions between ``pasta'' phases in
  dense matter},}\ }\href {\doibase 10.1103/PhysRevLett.94.031101} {\bibfield
  {journal} {\bibinfo  {journal} {Phys. Rev. Lett.}\ }\textbf {\bibinfo
  {volume} {94}},\ \bibinfo {pages} {031101}}\BibitemShut {NoStop}%
\bibitem [{\citenamefont {Watanabe}\ \emph {et~al.}(2002)\citenamefont
  {Watanabe}, \citenamefont {Sato}, \citenamefont {Yasuoka},\ and\
  \citenamefont {Ebisuzaki}}]{PhysRevC.66.012801}%
  \BibitemOpen
  \bibfield  {author} {\bibinfo {author} {\bibnamefont {Watanabe},
  \bibfnamefont {Gentaro}}, \bibinfo {author} {\bibfnamefont {Katsuhiko}\
  \bibnamefont {Sato}}, \bibinfo {author} {\bibfnamefont {Kenji}\ \bibnamefont
  {Yasuoka}}, \ and\ \bibinfo {author} {\bibfnamefont {Toshikazu}\ \bibnamefont
  {Ebisuzaki}}} (\bibinfo {year} {2002}),\ \bibfield  {title} {\enquote
  {\bibinfo {title} {Microscopic study of slablike and rodlike nuclei: Quantum
  molecular dynamics approach},}\ }\href {\doibase 10.1103/PhysRevC.66.012801}
  {\bibfield  {journal} {\bibinfo  {journal} {Phys. Rev. C}\ }\textbf {\bibinfo
  {volume} {66}},\ \bibinfo {pages} {012801}}\BibitemShut {NoStop}%
\bibitem [{\citenamefont {Watanabe}\ \emph {et~al.}(2003)\citenamefont
  {Watanabe}, \citenamefont {Sato}, \citenamefont {Yasuoka},\ and\
  \citenamefont {Ebisuzaki}}]{PhysRevC.68.035806}%
  \BibitemOpen
  \bibfield  {author} {\bibinfo {author} {\bibnamefont {Watanabe},
  \bibfnamefont {Gentaro}}, \bibinfo {author} {\bibfnamefont {Katsuhiko}\
  \bibnamefont {Sato}}, \bibinfo {author} {\bibfnamefont {Kenji}\ \bibnamefont
  {Yasuoka}}, \ and\ \bibinfo {author} {\bibfnamefont {Toshikazu}\ \bibnamefont
  {Ebisuzaki}}} (\bibinfo {year} {2003}),\ \bibfield  {title} {\enquote
  {\bibinfo {title} {Structure of cold nuclear matter at subnuclear densities
  by quantum molecular dynamics},}\ }\href {\doibase
  10.1103/PhysRevC.68.035806} {\bibfield  {journal} {\bibinfo  {journal} {Phys.
  Rev. C}\ }\textbf {\bibinfo {volume} {68}},\ \bibinfo {pages}
  {035806}}\BibitemShut {NoStop}%
\bibitem [{\citenamefont {Watanabe}\ \emph {et~al.}(2004)\citenamefont
  {Watanabe}, \citenamefont {Sato}, \citenamefont {Yasuoka},\ and\
  \citenamefont {Ebisuzaki}}]{PhysRevC.69.055805}%
  \BibitemOpen
  \bibfield  {author} {\bibinfo {author} {\bibnamefont {Watanabe},
  \bibfnamefont {Gentaro}}, \bibinfo {author} {\bibfnamefont {Katsuhiko}\
  \bibnamefont {Sato}}, \bibinfo {author} {\bibfnamefont {Kenji}\ \bibnamefont
  {Yasuoka}}, \ and\ \bibinfo {author} {\bibfnamefont {Toshikazu}\ \bibnamefont
  {Ebisuzaki}}} (\bibinfo {year} {2004}),\ \bibfield  {title} {\enquote
  {\bibinfo {title} {Phases of hot nuclear matter at subnuclear densities},}\
  }\href {\doibase 10.1103/PhysRevC.69.055805} {\bibfield  {journal} {\bibinfo
  {journal} {Phys. Rev. C}\ }\textbf {\bibinfo {volume} {69}},\ \bibinfo
  {pages} {055805}}\BibitemShut {NoStop}%
\bibitem [{\citenamefont {Watanabe}\ and\ \citenamefont
  {Sonoda}(2007)}]{Watanabe:2005qt}%
  \BibitemOpen
  \bibfield  {author} {\bibinfo {author} {\bibnamefont {Watanabe},
  \bibfnamefont {Gentaro}}, \ and\ \bibinfo {author} {\bibfnamefont {Hidetaka}\
  \bibnamefont {Sonoda}}} (\bibinfo {year} {2007}),\ \bibfield  {title}
  {\enquote {\bibinfo {title} {{Dynamical simulation of nuclear 'pasta': soft
  condensed matter in dense stars}},}\ }\href@noop {} {\bibfield  {journal}
  {\bibinfo  {journal} {Soft Condensed Matter: New Research}\ }}\Eprint
  {http://arxiv.org/abs/cond-mat/0502515} {arXiv:cond-mat/0502515
  [cond-mat.soft]} \BibitemShut {NoStop}%
\bibitem [{\citenamefont {Watanabe}\ \emph {et~al.}(2009)\citenamefont
  {Watanabe}, \citenamefont {Sonoda}, \citenamefont {Maruyama}, \citenamefont
  {Sato}, \citenamefont {Yasuoka},\ and\ \citenamefont
  {Ebisuzaki}}]{PhysRevLett.103.121101}%
  \BibitemOpen
  \bibfield  {author} {\bibinfo {author} {\bibnamefont {Watanabe},
  \bibfnamefont {Gentaro}}, \bibinfo {author} {\bibfnamefont {Hidetaka}\
  \bibnamefont {Sonoda}}, \bibinfo {author} {\bibfnamefont {Toshiki}\
  \bibnamefont {Maruyama}}, \bibinfo {author} {\bibfnamefont {Katsuhiko}\
  \bibnamefont {Sato}}, \bibinfo {author} {\bibfnamefont {Kenji}\ \bibnamefont
  {Yasuoka}}, \ and\ \bibinfo {author} {\bibfnamefont {Toshikazu}\ \bibnamefont
  {Ebisuzaki}}} (\bibinfo {year} {2009}),\ \bibfield  {title} {\enquote
  {\bibinfo {title} {Formation of nuclear ``pasta'' in supernovae},}\ }\href
  {\doibase 10.1103/PhysRevLett.103.121101} {\bibfield  {journal} {\bibinfo
  {journal} {Phys. Rev. Lett.}\ }\textbf {\bibinfo {volume} {103}},\ \bibinfo
  {pages} {121101}}\BibitemShut {NoStop}%
\bibitem [{\citenamefont {Wijnands}\ and\ \citenamefont
  {et~al.}(2004)}]{crustcooling}%
  \BibitemOpen
  \bibfield  {author} {\bibinfo {author} {\bibnamefont {Wijnands},
  \bibfnamefont {R}}, \ and\ \bibinfo {author} {\bibnamefont {et~al.}}}
  (\bibinfo {year} {2004}),\ \href@noop {} {\bibinfo  {journal}
  {astro-ph/0405089}\ }\BibitemShut {NoStop}%
\bibitem [{\citenamefont {Williams}\ and\ \citenamefont
  {Koonin}(1985)}]{WILLIAMS1985844}%
  \BibitemOpen
\bibfield  {journal} {  }\bibfield  {author} {\bibinfo {author} {\bibnamefont
  {Williams}, \bibfnamefont {RD}}, \ and\ \bibinfo {author} {\bibfnamefont
  {S.E.}\ \bibnamefont {Koonin}}} (\bibinfo {year} {1985}),\ \bibfield  {title}
  {\enquote {\bibinfo {title} {Sub-saturation phases of nuclear matter},}\
  }\href {\doibase http://dx.doi.org/10.1016/0375-9474(85)90191-5} {\bibfield
  {journal} {\bibinfo  {journal} {Nuclear Physics A}\ }\textbf {\bibinfo
  {volume} {435}}~(\bibinfo {number} {3}),\ \bibinfo {pages} {844 --
  858}}\BibitemShut {NoStop}%
\bibitem [{\citenamefont {Winget}\ and\ \citenamefont {et~al}(2009)}]{winget}%
  \BibitemOpen
  \bibfield  {author} {\bibinfo {author} {\bibnamefont {Winget}, \bibfnamefont
  {D~E}}, \ and\ \bibinfo {author} {\bibnamefont {et~al}}} (\bibinfo {year}
  {2009}),\ \href@noop {} {\bibfield  {journal} {\bibinfo  {journal} {ApJ.}\
  }\textbf {\bibinfo {volume} {693}},\ \bibinfo {pages} {L6}}\BibitemShut
  {NoStop}%
\bibitem [{\citenamefont {Woosley}\ \emph {et~al.}(2004)\citenamefont
  {Woosley}, \citenamefont {Heger}, \citenamefont {Cumming}, \citenamefont
  {Hoffman}, \citenamefont {Pruet}, \citenamefont {Rauscher}, \citenamefont
  {Fisker}, \citenamefont {Schatz}, \citenamefont {Brown},\ and\ \citenamefont
  {Wiescher}}]{rpash2}%
  \BibitemOpen
  \bibfield  {author} {\bibinfo {author} {\bibnamefont {Woosley}, \bibfnamefont
  {S~E}}, \bibinfo {author} {\bibfnamefont {A.}~\bibnamefont {Heger}}, \bibinfo
  {author} {\bibfnamefont {A.}~\bibnamefont {Cumming}}, \bibinfo {author}
  {\bibfnamefont {R.~D.}\ \bibnamefont {Hoffman}}, \bibinfo {author}
  {\bibfnamefont {J.}~\bibnamefont {Pruet}}, \bibinfo {author} {\bibfnamefont
  {T.}~\bibnamefont {Rauscher}}, \bibinfo {author} {\bibfnamefont {J.~L.}\
  \bibnamefont {Fisker}}, \bibinfo {author} {\bibfnamefont {H.}~\bibnamefont
  {Schatz}}, \bibinfo {author} {\bibfnamefont {B.~A.}\ \bibnamefont {Brown}}, \
  and\ \bibinfo {author} {\bibfnamefont {M.}~\bibnamefont {Wiescher}}}
  (\bibinfo {year} {2004}),\ \href@noop {} {\bibfield  {journal} {\bibinfo
  {journal} {ApJ Supp.}\ }\textbf {\bibinfo {volume} {151}},\ \bibinfo {pages}
  {75}}\BibitemShut {NoStop}%
\end{thebibliography}

%

\end{document}